\documentclass{aa}

\usepackage{graphicx}   
\usepackage[colorlinks=true,linkcolor=blue,citecolor=blue, urlcolor=blue]{hyperref}
\usepackage{txfonts}
\usepackage{enumitem}
\usepackage{subcaption}

\newcommand{\nimg}{n_{\rm img}}
\newcommand{\Dori}{D_{\rm ori}}
\newcommand{\Denh}{D_{\rm shp}}

\newcommand{\Denhm}{D_{\rm shp}^{\rm mask}}

\newcommand{\Gav}{G_{\rm av}}
\newcommand{\Chitah}{{\sc Chitah}}

\newcommand{\bd}{\begin{displaymath}}
\newcommand{\ed}{\end{displaymath}}
\newcommand{\be}{\begin{equation}}
\newcommand{\ee}{\end{equation}}
\newcommand{\beaa}{\begin{eqnarray*}}
\newcommand{\eeaa}{\end{eqnarray*}}
\newcommand{\bea}{\begin{eqnarray}}
\newcommand{\eea}{\end{eqnarray}}

\newcommand{\sref}[1]{Sect.~\ref{#1}}
\newcommand{\ssref}[1]{Sects.~\ref{#1}}
\newcommand{\aref}[1]{Appendix~\ref{#1}}
\newcommand{\fref}[1]{Fig.~\ref{#1}}
\newcommand{\fsref}[1]{Figs.~\ref{#1}}
\newcommand{\tref}[1]{Table~\ref{#1}}

\newcommand{\eref}[1]{Eq.~(\ref{#1})}

\begin{document} 


\title{Discovery of strongly lensed quasars in the Ultraviolet Near Infrared Optical Northern Survey (UNIONS)}
\titlerunning{Lensed QSOs in UNIONS}

\author{
J.~H.~H.~Chan\inst{\ref{epfl}}\and
C.~Lemon\inst{\ref{epfl}}\and
F.~Courbin\inst{\ref{epfl}}\and
R.~Gavazzi\inst{\ref{iap},\ref{ioa}}\and
B.~Cl\'ement\inst{\ref{epfl}}\and
M.~Millon\inst{\ref{epfl}}\and
E.~Paic\inst{\ref{epfl}}\and
K.~Rojas\inst{\ref{epfl}}\and
E.~Savary\inst{\ref{epfl}}\and
G.~Vernardos\inst{\ref{epfl}}\and
J.-C.~Cuillandre\inst{\ref{aim}}\and 
S.~Fabbro\inst{\ref{nrc}}\and
S.~Gwyn\inst{\ref{nrc}}\and
M.~J.~Hudson\inst{\ref{waterloo},\ref{physwaterloo},\ref{pi}}\and
M.~Kilbinger\inst{\ref{aim}}\and
A.~McConnachie\inst{\ref{nrc}}
}

\institute{
Institute of Physics, Laboratory of Astrophysique, \'Ecole Polytechnique F\'ed\'erale de Lausanne (EPFL), Observatoire de Sauverny, 1290 Versoix, Switzerland 
\label{epfl}
\newline
\email{hung-hsu.chan@epfl.ch}
\and
Institut d'Astrophysique de Paris, UMR7095 CNRS \& Sorbonne Universit\'e, 98bis Bd Arago, 75014 Paris, France \label{iap} 
\and
Institute of Astronomy, University of Cambridge, Madingley Road, Cambridge CB30HA, UK
\label{ioa}
\and
AIM, CEA, CNRS, Universit\'e Paris-Saclay, Universit\'e de Paris, F-91191 Gif-sur-Yvette, France 
\label{aim}
\and
NRC Herzberg Astronomy \& Astrophysics, 5071 West Saanich Road, Victoria, BC, V9E 2E7, Canada
\label{nrc}
\and
Waterloo Centre for Astrophysics, University of Waterloo, 200, University Ave W, Waterloo, ON N2L 3G1 , Canada
\label{waterloo}
\and
Department of Physics and Astronomy, University of Waterloo, Waterloo, ON, N2L 3G1, Canada          
\label{physwaterloo}
\and
Perimeter Institute for Theoretical Physics, 31 Caroline St N, Waterloo, ON N2L 2Y5, Canada
\label{pi}
}

\date{\today}

\abstract{
We report the discovery of five new doubly imaged lensed quasars from the first 2\,500 square degrees of the ongoing Canada-France Imaging Survey (CFIS), which is a component of the Ultraviolet Near Infrared Optical Northern Survey (UNIONS). The systems are preselected in the initial catalogues of either {\it Gaia} pairs or MILLIQUAS quasars. We then take advantage of the deep, 0.6\arcsec median-seeing $r$-band imaging of CFIS to confirm the presence of multiple point sources with similar colour of $u-r$ via convolution of the Laplacian of the point spread function. Requiring point sources of similar colour and with flux ratios of less than 2.5~mag in $r$-band, we reduce the number of candidates from 256\,314 to 7\,815. After visual inspection, we obtain 30 high-grade candidates, and prioritise a spectroscopic follow-up analysis for those showing signs of a lensing galaxy upon subtraction of the point sources. We obtain long-slit spectra for 18 candidates with ALFOSC on the 2.56-m Nordic Optical Telescope (NOT), confirming five new doubly lensed quasars with $1.21<z<3.36$ and angular separations from 0.8\arcsec\ to 2.5\arcsec. One additional system is a probable lensed quasar based on the CFIS imaging and existing SDSS spectrum. We further classify six objects as nearly identical quasars, that is, possible lenses but without the detection of a lensing galaxy. Given our recovery rate ($83\%$) of existing optically bright lenses within the CFIS footprint, we expect that a similar strategy, coupled with $u-r$ colour-selection from CFIS alone, will provide an efficient and complete discovery of small-separation lensed quasars of source redshifts below $z=2.7$ within the CFIS $r$-band magnitude limit of 24.1~mag.
}

\maketitle


\section{Introduction} 
\label{sec:intro}

Gravitationally lensed quasars provide a powerful means to study both galaxy evolution and cosmology, and appear when a foreground lensing galaxy deflects the light from a distant quasar source. The configuration of lensed images enables us to study the mass structures and substructures of lensing galaxies and to further probe galaxy evolution in the Universe \citep{SuyuEtal12,Dalal&Kochanek02,VegettiEtal12,NierenbergEtal17,GilmanEtal19}. The time delays between image pairs can be used to infer the Hubble constant, $H_0$, a crucial cosmological parameter in determining the size, age, and critical density of the Universe \citep{Refsdal64,ChenEtal19,WongEtal19}. Given the current tension between $H_0$ measurements from the early and late Universe, independent measurements from lensed quasars are vital \citep{VerdeEtal19}.

There are currently around 200 known gravitationally lensed quasars. The Cosmic Lens All-Sky Survey \citep[CLASS;][]{MyersEtal03} provided the largest statistical sample of radio-loud lensed quasars by identifying the flat-spectrum radio sources with multiple lensed images. From optical data, the SDSS Quasar Lens Search discovered 62 lenses based on both morphological and colour selection of spectroscopically confirmed quasars \citep{OguriEtal06,OguriEtal08,OguriEtal12,InadaEtal08,InadaEtal10,InadaEtal12}. \cite{JacksonEtal12} extended the SQLS sample by using the better image quality from the UKIRT Infrared Deep Sky Survey (UKIDSS). Given the increasing depth and image quality of ongoing surveys, it is now possible to identify lensed quasars from imaging alone, with no spectroscopic pre-selection.

\cite{ChanEtal15} demonstrated such a search using \Chitah\ -- an automated pixel- and mass-modelling code for finding lensed quasars in the Hyper-Suprime Cam (HSC) Survey \citep{AiharaEtal18}. Other image-based lens-finding strategies have relied on machine learning, ring-finders, and citizen science \citep{SonnenfeldEtal18,SonnenfeldEtal19,WongEtal18,ChanEtal20,SonnenfeldEtal20,JaelaniEtal20,JaelaniEtal21}, however these have mainly had success for lensed galaxies where the extended lensed arcs are distinguishable from other astrophysical objects. In the case of lensed quasars, often just two point sources are present, which outshine the light from the lensing galaxy and are outnumbered by visually similar contaminants like binary stars or quasar--star projections. \citet{LemonEtal18,LemonEtal19} used simultaneous optical--infrared extracted colours combined with astrometric measurements from {\it Gaia} to avoid such contaminants. Thanks to its exceptional spatial resolution, the catalogued image positions in {\it Gaia} have been used to identify new lenses \citep{DucourantEtal18a,DucourantEtal18b,Krone-MartinsEtal18}.

Here, we present a search for lensed quasars in the Canada-France Imaging Survey \citep[CFIS;][]{IbataEtal17}. We analyse CFIS Data Release 2 (DR2) data, covering $2\,500\deg^2$. While the excellent image quality and depth of the $r$-band allow for detection of faint quasar images and lensing galaxies, the $u$-band provides an important colour indicator for removal of many common contaminants. Despite previous spectroscopic searches in this area of the sky, we expect many lenses to still be left undiscovered either due to lack of depth and/or stringent pre-selection of previous search techniques. The CFIS data, thanks to their depth and seeing, also consist in a major advantage to find these missing lenses. 

This paper is organised as follows. The descriptions of the CFIS pixel data and the quasar catalogue selection are provided in \sref{sec:data}. We describe the automated part of our lens search method in \sref{sec:method}, and the visual inspection stage in \sref{sec:inspect}. Prioritisation of high-grade systems for spectroscopy is outlined in \sref{sec:light}, with the results of follow-up spectroscopy presented in \sref{sec:not}. We present our results and discuss the individual systems in \sref{sec:result}. In \sref{sec:conclusion} we conclude our findings. 
All images are oriented with north up and east to the left. Optical magnitudes quoted in this paper are in the AB system, and infrared magnitudes in the Vega system. When required, a flat cosmology with $H_0=70~{\rm km~s^{-1}~Mpc^{-1}}$, $\Omega_m=0.3$ and $\Omega_\Lambda=0.7$ is used.

\section{Data}
\label{sec:data}

We first describe the available CFIS imaging data (\sref{subsec:cfis}), and then the initial quasar catalogues in which we perform our selection (\sref{subsec:catalog}).

\subsection{CFIS imaging data} 
\label{subsec:cfis}

The Canada-France Imaging Survey (CFIS) is a component of the Ultraviolet Near Infrared Optical Northern Survey (UNIONS) project, which is a collaboration of wide-field imaging surveys of the northern hemisphere. UNIONS consists of CFIS, conducted at the $3.6$-m Canada-France-Hawaii Telescope (CFHT) on Maunakea in Hawaii, members of the Pan-STARRS team, and the Wide Imaging with Subaru Hyper Suprime-Cam of the Euclid Sky (WISHES) team. CFHT/CFIS is obtaining deep $u$- and $r$-bands; Pan-STARRS is obtaining deep $i$-band and moderate-deep $z$-band imaging; and Subaru/WISHES is obtaining deep $z$-band imaging. These independent efforts are, in part, working towards securing optical imaging to complement the Euclid space mission, although UNIONS is a separate collaboration aimed at maximizing the science return of these large and deep surveys of the northern sky. \cite{IbataEtal17}, \cite{FantinEtal19}, and \cite{GuinotEtal21} provided an additional illustration of CFIS imaging data. This CFIS component is mapping the northern sky in the $u$-band (CFIS-$u$, covering $8\,000~\deg^2$) and $r$-band (CFIS-$r$, covering $4\,800~\deg^2$). For the present work, we use all available data from CFIS DR2 with both $u$- and $r$-band coverage, i.e. $\sim 2\,500~\deg^2$ \citep[see fig. 1 in][]{SavaryEtal21}. The full survey is expected to finish in 2023. CFIS-$r$ provides exquisite image quality with median seeing of $\sim0.6\arcsec$ to a depth of $24.1$ ($10\sigma$, 2\arcsec diameter aperture), and CFIS-$u$ has median seeing of $\sim0.8\arcsec$ to a depth of $23.6$. 

For all candidates, we generate image cutouts of $8.2\arcsec\times8.2\arcsec$ (44 pixels on-a-side). We also produce models of the point spread function (PSF) and its spatial variations across co-added images which have been reduced, processed, and calibrated at CADC using an improved version of the MegaPipe pipeline \citep{Gwyn08,Gwyn09}. The model PSF is exploited with PSFEx \citep{Bertin13} from the stacked image, and each image of the local PSF is oversampled by a factor of two. We notice that the optimal PSF model should be obtained from the individual exposure images, although the imperfect PSFs only marginally affect the detection of lensed quasars. More accurate and precise PSF models are beyond the scope of this work. Weight images along with other data quality diagnostics are also produced for each candidate in each of the available bands. The pixel scale of co-added images is $0.186\arcsec$.

\subsection{Quasar catalogues} 
\label{subsec:catalog}

The goal of this work is to find convincing lensed quasar candidates within the CFIS footprint, while minimising the manual inspection time. We therefore choose to start our search from catalogues of systems likely to contain at least one quasar. In future searches, we expect to be able to use the CFIS data themselves as a way to select quasars, because the $u$ and $r$-bands provide an efficient way to select quasars as objects with a $u$-band excess \citep[see e.g.][]{Nakoneczny2019}.

The Million Quasars Catalog v6.4 \citep[hereafter MILLIQUAS;][]{Flesch19} provides a compilation of both confirmed and candidate quasars. It is composed of all spectroscopically reported quasars, radio and X-ray detections, and probable quasars based on AllWISE colours and photometry \citep{SecrestEtal15}. Restricting the original catalogue of 1\,980\,903 objects to those with CFIS-$u$ and -$r$ data leaves 253\,299 systems. 

While around $30\%$ of the MILLIQUAS catalogue comprises quasar candidates selected from infrared colours \citep{SecrestEtal15}, the colour-selection targets isolated quasars. We expect that many lensed quasars will have bluer infrared colours either due to blending with the lensing galaxy, or due to having high-redshift ($z\gtrsim3$) sources. We therefore choose to augment the MILLIQUAS catalogue with a selection based on a bluer {\it WISE} limit of $W1-W2>0.5$. To minimise contaminants, we cross-match these {\it WISE} detections (requiring $W1<16.5$) to \textit{Gaia} detections, keeping only those systems with two \textit{Gaia} detections within $3\arcsec$ of the AllWISE catalogue detection \citep{SecrestEtal15,gaiadr2}. Finally, we impose a cut on the proper motion significance (PMSIG) of less than $15\sigma$ for each \textit{Gaia} detection, as defined in \citet{LemonEtal19}. This leaves 9\,244 systems within the CFIS DR2 footprint, of which only 3\,015 currently have both CFIS-$u$ and -$r$ imaging. 
Including 253\,299 systems in MILLIQUAS, we have 256\,314 pre-selected quasar candidates in total to be classified as lensed quasar candidates.

\section{Classification method} 
\label{sec:method}

Our search algorithm relies upon identifying systems with multiple, similar-colour point sources, as is expected in lensed quasars. This straightforward approach can be effective when applied to systems known to contain at least one quasar point source, because in the case of a quasar+star projection, we expect different $u-r$ colours \citep[e.g. see Fig. 1 in][]{FinlatorEtal20}. This separation of stars and quasars in colour-space is due to the $u$-band excess of quasars at redshifts $z \lesssim2.7$. At higher redshifts, the Lyman-alpha forest is redshifted into the $u$-band and we expect a higher rate of contamination from quasar+star projections. We note that the following algorithm is applicable to all imaging datasets with two or more bands; however we use CFIS as a testbed given its depth, excellent image quality, and the aforementioned use of the $u$-band.
\begin{figure}[h!]
\centering
\includegraphics[scale=0.6]{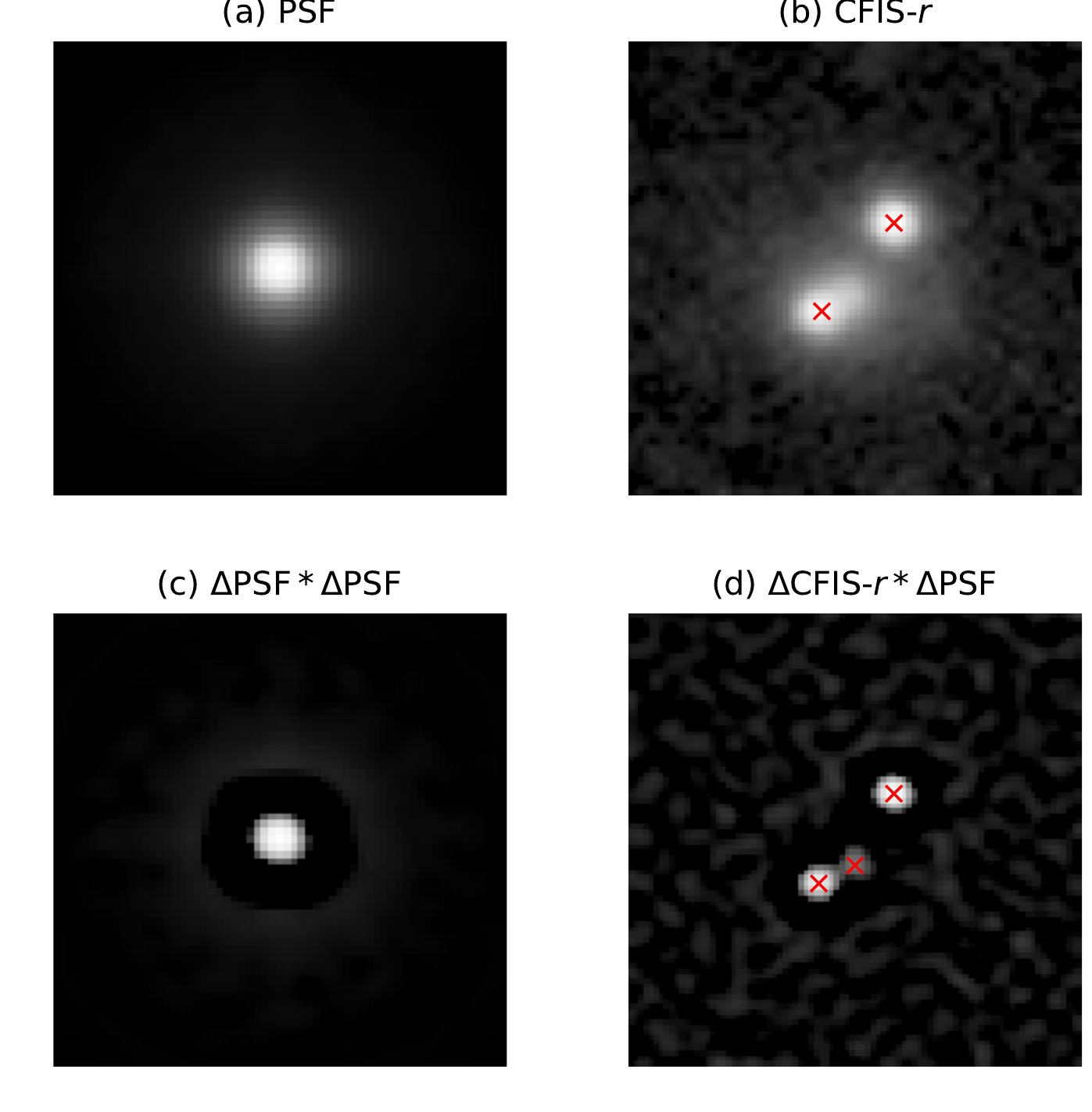}
\caption{
Point source detection with and without the convolution of Laplacian. (a) Measured PSF, (b) $r$-band data, (c) the sharpened image of PSF, and (d) the sharpened image of $r$-band data. The sharpened images are obtained by convolving the Laplacian of the image with the Laplacian of the corresponding PSF (see \eref{eqn:enhanced}). We replace the negative values with zeros in (c) and (d) due to the byproduct of this convolution. The red crosses in (b) and (d) indicate the point source detection using \texttt{DAOStarFinder} with 2D Gaussian kernels fitted from (a) and (c), respectively. The sharpened image (d) with improved dynamical range allows us to detect blended images better as shown in the red central cross.
}
\label{fig:enhanced}
\end{figure}
The search algorithm is as follows:
\begin{enumerate}
    \item {\it Sharpening the point sources:}
    Before comparing colours of multiple point sources, we require reliable identification of point sources in the ground-based imaging, where nearby point sources are often overlapping and blended. Conventional extraction algorithms are limited by this blending, and adjusting the deblending threshold parameters leads to multiple spurious detections from extended sources. However, we can identify positions of likely point sources in each of our candidate cutouts because we have the expected two-dimensional PSF profile at the location of the system. To achieve this, we measure the second derivative of the observed surface brightness and compare it with the second derivative of the PSF \citep[See Fig. 7 in][]{CantaleEtal16}.
    In practice, we calculate the convolution of the Laplacian of the original image, $\Dori$, and the Laplacian of its corresponding PSF to obtain the sharpened image, $\Denh$, as:
    \be
    \Denh=\Delta\Dori * \Delta\text{PSF},
    \label{eqn:enhanced}
    \ee
    where $\Delta$ is the Laplacian operator and $*$ is the convolution operator. The effect of this convolution is to reduce the size of the PSF FWHM by around 25\%, allowing more efficient detection of small separation point sources. However, a by-product of this convolution on any PSF is a negative annulus around the sharpened PSF. We replace any negative values with zeros in $\Denh$, providing the final image, denoted as $\Denhm$, for the subsequent point source detection. In \fref{fig:enhanced} we illustrate this sharpening on a known lens. As we have imaging data in two filters well-separated in colour, we can separate the lens and the lensed images based on the colour measurements, as described below.
  
    \item {\it Identifying the point sources:}
    After obtaining the sharpened images, including their sharpened PSFs, we use the python module \texttt{DAOStarFinder}\footnote{\url{https://photutils.readthedocs.io/en/stable/api/photutils.detection.DAOStarFinder.html}} for locating the possible point sources; see \fsref{fig:enhanced} and \ref{fig:chitah}. DAOFIND, used in \texttt{DAOStarFinder}, searches images for local maxima with a peak amplitude greater than a given threshold and with a size and shape similar to the defined 2D Gaussian kernel \citep{Stetson87}. During this process, we supply the 2D Gaussian fit of the local PSF to \texttt{DAOStarFinder}\footnote{We empirically choose $\texttt{threshold}=30.0\cdot\texttt{std}$, where $\texttt{std}$ is estimated using $\texttt{sigma\_clipped\_stats}$. The other parameters of \texttt{DAOStarfinder} are set as default.}. The result of this process is to return positions of identified point sources. In this step, we do not consider the flux measurements even though they are part of the output, as the sharpened images and PSFs create a bias in the flux estimation. We illustrate this in \fref{fig:enhanced} (b) and (d).

    \item {\it Grouping the point sources by position:}
    The identification of point sources is run separately on each band, and can lead to different detections per system. This can be due to low signal-to-noise ratio in one band, differing PSF widths, or a very red or blue spectral energy distribution (SED). We collect all detections across both bands, and group any double detections into one detection when their distances are less than 1.5 pixels in the oversampled image (i.e. $0.14\arcsec$), as shown in \fref{fig:chitah}. The final position is taken as the mean position of each group.
    
    \item {\it Removing systems with one or no identified point sources:}
    We remove any system with one or no point sources detected in the preceding step, as it is unlikely to contain the multiple point sources seen in a lensed quasar.
    
    \item {\it Measuring the fluxes of point sources:}
    To extract the colours of each component, we first estimate the flux in each band by performing a least-squares fit of PSFs placed at their derived positions from step 3. The relevant $\chi^2$ is:
    \be
    \chi^2=\sum_{i,j}\frac{\left[ \sum\limits_{k=1}^{n}P_k(i,j)-\Dori(i,j) \right]^2}{var(i,j)},
    \ee
    where $i=1...N_x$ and $j=1...N_y$ are the pixel indices of the image cutout of dimensions $N_x\times N_y$, $P_k(i,j)$ is the $k$-th scaled PSF placed at the measured position, and $var(i,j)$ is the pixel variance of $\Dori(i,j)$. As we oversample our cutouts by a factor of two, $N_x=N_y=88$. We note that, if detected, the fluxes of lensing galaxies are likely underestimated, but this is partly cancelled out when deriving the colour of the extended component.
    
    \item {\it Removing the faint point sources:}
    To avoid the false detection on the noise peaks, we abandon the point sources beyond the observation depth of each tile in both bands. In cases where we miss faint lensed images, we extend the observation limit in CFIS to about 0.5~mag larger than the depth of $10\sigma$ point source.
    
    \item {\it Grouping the blue point sources:}
    A main contaminant of our sample is single quasars with a nearby or coincident star or galaxy. Therefore we aim at separating the quasars from other objects, expecting quasars to be the bluest objects in any cutout. We first identify the bluest point source, i.e. the smallest $u-r$ value, denoted as $u_{\rm b}-r_{\rm b}$, and collect any other detections with a $u-r$ value within 1~mag of this bluest detection, i.e. ${|(u_k-r_k)-(u_{\rm b}-r_{\rm b})|<1}$~mag. We assign these detections, along with the original, as the blue group and any remaining point sources are assembled as the red group. As shown in \fref{fig:chitah} (a) and (b) as an example, circles 1 and 2 (in blue and orange respectively) are classified in the blue group and diamond 3 (in green) is classified in the red group. We note that sometimes an object can be composed of only one group, particularly when no galaxies are present. Also the blue group may be arbitrarily red, for example when a quasar is dusty or at high redshift. 
    
    \item {\it Selecting systems with multiple blue sources:}
    As typical lensed quasar systems contain two or four images, we classify systems with  $2\leq\nimg\leq5$ as lens candidates, where $\nimg$ is the number of detections in the blue group. The upper limit of $\nimg=5$ is empirically chosen to reduce contamination from dense star-forming galaxies or high-density stellar fields.
    
\end{enumerate}
\begin{figure}[h!]
\centering
\includegraphics[scale=0.6]{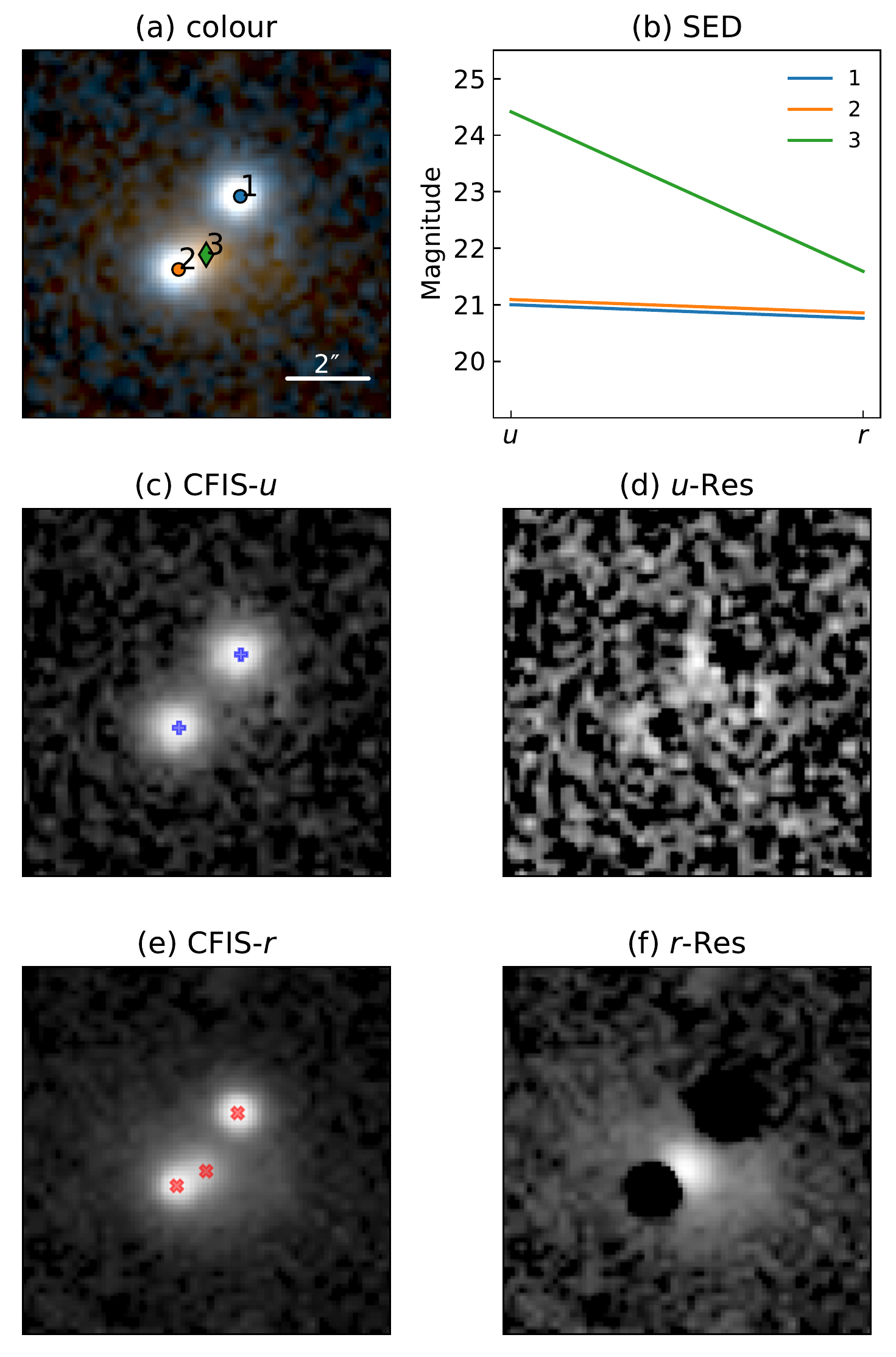}
\caption{
Output of the classification method on the known lens SDSSJ1537+3014. (a) CFIS composite image made from the $u$ and $r$ bands, (b) SED of the detected point sources, (c) $u$-band data, (d) $u$-band residuals upon subtraction of blue group point sources, (e) $r$-band data, and (f) $r$-band residuals upon subtraction of the blue point sources. Circles and diamonds in (a) show the positions of the blue and red point sources, respectively, with single-band detections overlaid in panels (c) and (e).
}
\label{fig:chitah}
\end{figure}
This method was developed alongside simultaneous tests on CFIS data of known lenses and contaminant systems. In particular, we used a set of 18 optically bright, known lensed quasars with CFIS $u$ and $r$ band imaging \citep[from the lens compilation of][]{LemonEtal19}\footnote{\url{https://research.ast.cam.ac.uk/lensedquasars/index.html}}. We recover 15 of these 18 lensed quasars, i.e. a recovery rate of $83\%$, and summarise these tests and the three failures in \aref{sec:known}. We now apply this automated search algorithm to the 256\,314 pre-selected quasar candidates in \sref{sec:data}, retaining 10\,914 systems, i.e. reducing the initial catalogue to 4.3\% of its original size. We compare the colours and flux ratios of this sample to the 15 recovered lenses in the CFIS footprint (see \fref{fig:select}). The colour difference is measured using the brightest two potential lensed images (i.e. from the bluest group following step 7). We note that there is a limit on the absolute colour difference, as imposed during our search algorithm (i.e. ${|(u_k-r_k)-(u_{\rm b}-r_{\rm b})|<1}$~mag).

\begin{figure}[h!]
\centering
\includegraphics[scale=0.65]{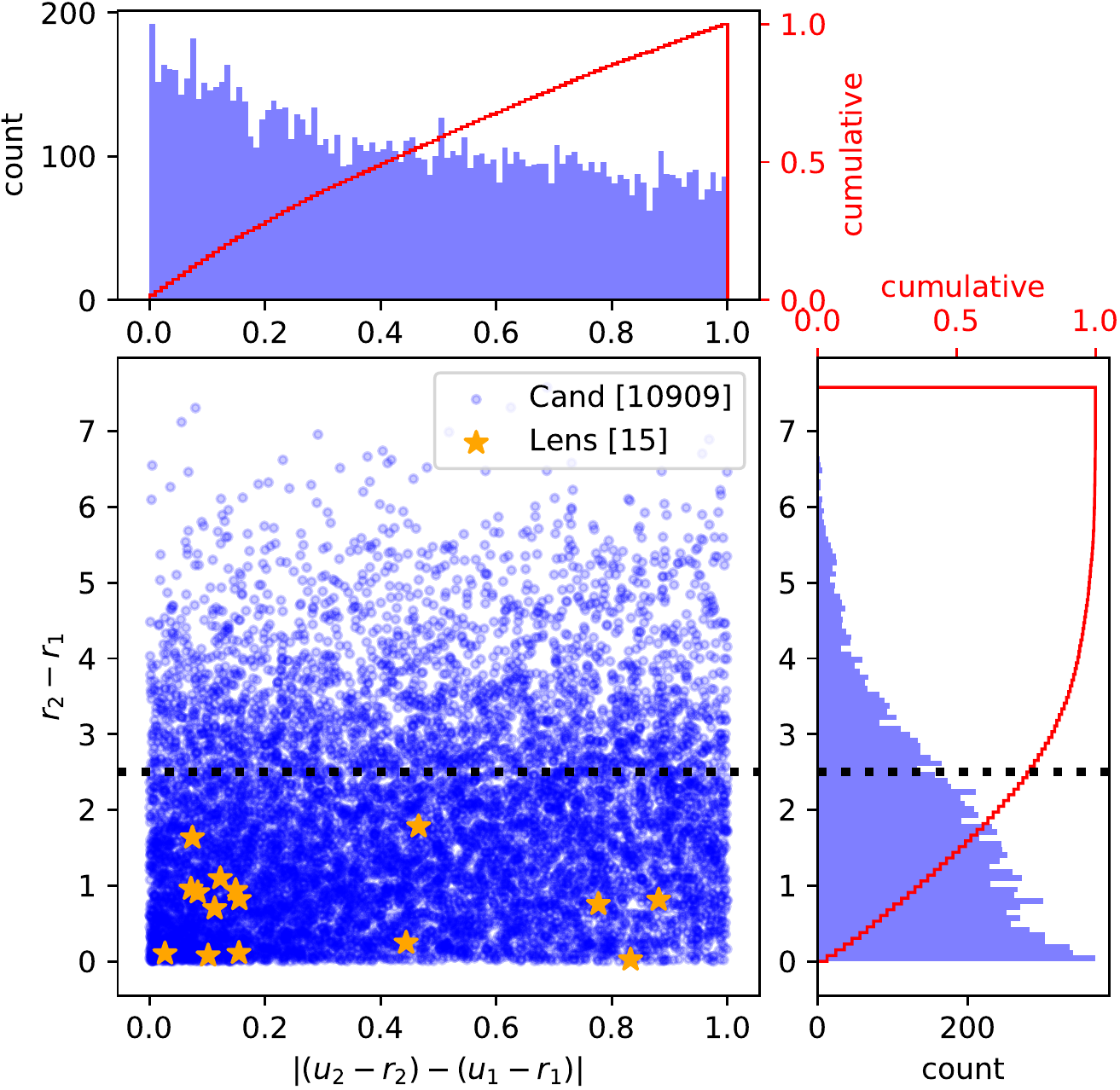}
\caption{
Flux-ratio against colour difference for systems passing the algorithm detailed in \sref{sec:method}, with known lenses overlaid as yellow stars. In the event of more than two detections, fluxes are based on the brightest detections within the blue group. The horizontal dashed line shows the flux ratio criterion we apply to further reduce the number of candidates.
}
\label{fig:select}
\end{figure}

As shown in \fref{fig:select}, there are many candidate systems with large flux ratios that are not compatible with lens models and not seen in known lenses. Inspecting several of these high-flux-ratio systems indeed shows that they are mostly associated to companions near bright stars, or star-forming galaxies. We therefore apply a cut in the flux ratio, at an absolute magnitude difference of 2.5~mag in $r$-band, reducing the number of candidates by $20\%$. We retain 7\,815 candidates for the next step of visual inspection.

\section{Visual inspection} 
\label{sec:inspect}

We propose two phases for the visual inspection, namely `rapid' and `refined', performed by two authors (JC and CL). During the phase of rapid inspection, we adopt a simple binary classification: `possible lens' and `non-lens'. For each candidate, we inspect an image of the system with the results of the analysis pipeline of \sref{sec:method}. An example of such an image is shown as \fref{fig:chitah}. We include the residual images after subtracting the point sources associated to the blue group in order to highlight any potentially faint compact lensing galaxies. The majority of the candidates are classed as non-lens for one or several of the following reasons: 
\begin{enumerate} 
    \item bad pixels in the cutout which lead to spurious point source detections from our pipeline;
    \item obvious extended galaxies or non-point sources, such as the host galaxy of a low-redshift AGN or a galaxy merger;
    \item noisy $u$-band data; 
    \item wide-separation point sources with no apparent lensing galaxy.
\end{enumerate}
\begin{figure}[h]
\centering
\includegraphics[scale=0.35]{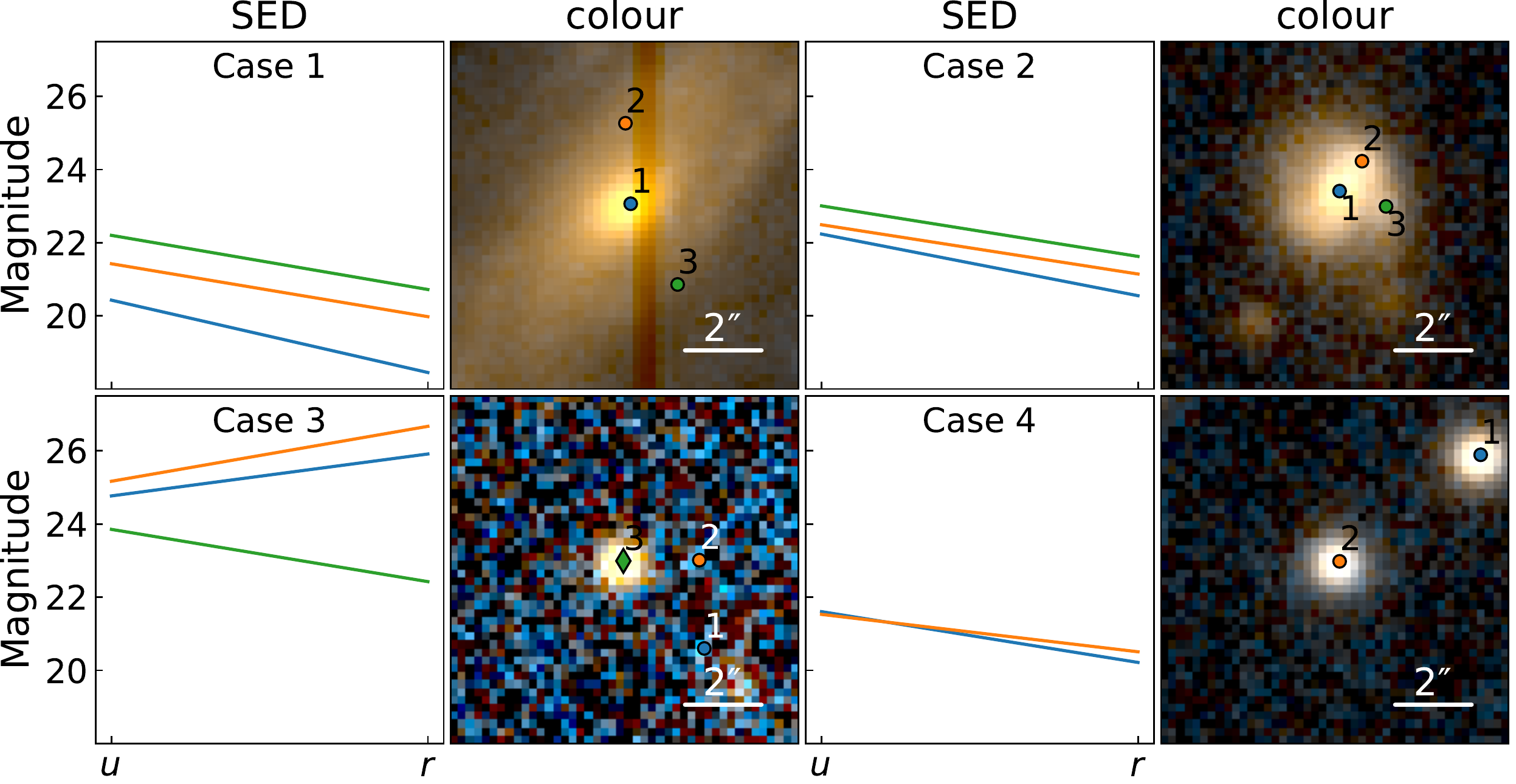}
\caption{
Examples of objects classified as non-lens during the rapid visual inspection because of bad pixels (case~1), mergers/low-redshift AGNs (case~2), noise peaks (case~3), or wide binaries (case~4).
}
\label{fig:false}
\end{figure}
Each case is illustrated in \fref{fig:false}. In hindsight, the third case could have been avoided by increasing the threshold during the classification method; however, these are easily removed during our rapid inspection phase. Of the 7\,815 inspected systems, CL/JC classifies 514/640 as `possible lens'. At this stage, we check that both inspectors recovered all of the 15 known optically bright lensed quasars in the sample (see \aref{sec:known} for the classification of known lenses by the selection method of \sref{sec:method}). Given the sufficient number for inspecting the candidates in detail, i.e the phase of refined inspection, we take the intersection of the two classifications as 239.

For each of these of 239 systems, we carefully inspect any existing data, including SDSS images and spectra, DECaLS images\footnote{\url{https://www.legacysurvey.org}} \citep{DeyEtal19}, unWISE images \citep{MeisnerEtal18}, and {\it Gaia} DR2 proper motions \citep[converted to the PMSIG parameter defined in][]{LemonEtal19}, and astrometric excess noise \citep[AEN;][]{gaiadr2}. A large PMSIG ($\gtrsim 10\sigma$) is a strong indicator that the system contains a star, and a high AEN ($\gtrsim 10$~mas) suggests that a candidate is possibly a star-forming galaxy \citep[see Figs. 1 and 2 in ][]{LemonEtal19}. We grade the candidates with values of between 0 and 10, where: 
\begin{itemize}[leftmargin=36pt]
    \item[10 - 9:] very likely to be a lens, requiring resolved spectroscopy for confirmation,
    \item[ 8 - 7:] blue point sources at modest separation with similar quasar-like colours and without an obvious lensing galaxy,
    \item[ 6 - 5:] same as the grades 8 - 7 but with slightly high PMSIG, AEN, and/or more different colours,
    \item[ 4 - 0:] a system that likely contains a star, star-forming galaxy, or a very low-redshift quasar.
\end{itemize}

\begin{figure*}[h]
\begin{subfigure}{\textwidth}
\centering
\includegraphics[scale=0.33 ]{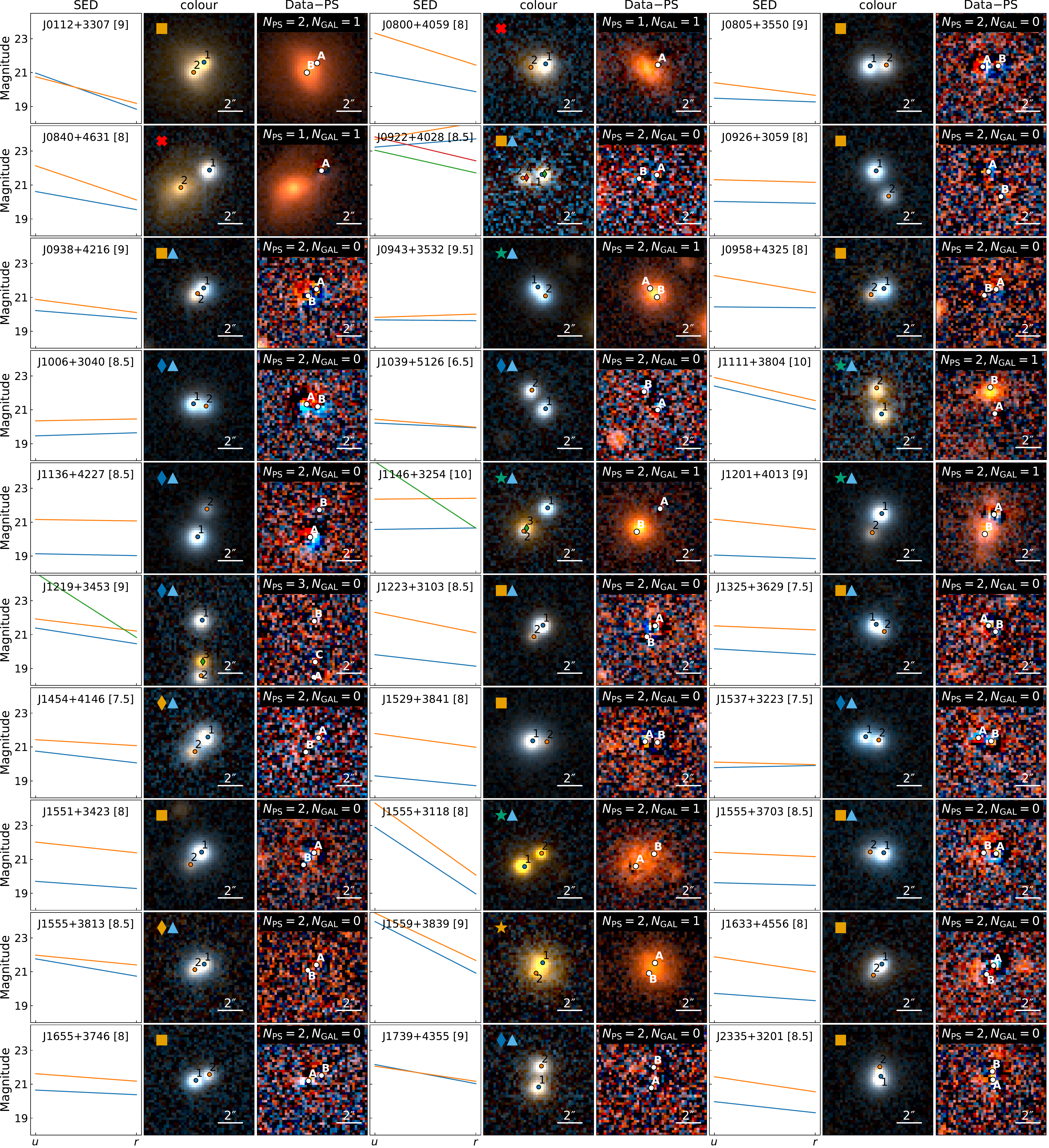}
\end{subfigure}
\begin{subfigure}{\textwidth}
\centering
\includegraphics[scale=0.33 ]{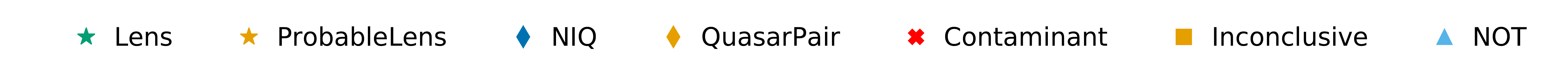}
\end{subfigure}
\caption{
Lens candidates with $\Gav\geq8$ or with follow-up spectroscopy with ALFOSC on the NOT. Each candidate contains `SED', `colour', and `Data$-$PS' from light modelling. The labels in `SED' and `colour' are the same as \fref{fig:chitah} (a) and (b). The number in square brackets indicates the grade of visual inspection $[\Gav]$. `Data$-$PS' illustrates the residual of point source subtraction. The numbers of point sources and galaxies for the fit are indicated at the the top of the panels as $N_{\rm PS}$ and $N_{\rm GAL}$, respectively. The results of light modelling for each component are listed in \tref{tab:cand_mag}. We label the confirmed lenses with green stars, the probable lenses with yellow stars, the NIQs (nearly identical quasars) with blue diamonds, the quasar pairs with yellow diamonds, the contaminants with red crosses, and the inconclusive systems with yellow squares. The systems with NOT spectra are labelled with cyan triangles.
}
\label{fig:cand_G8NOT}
\end{figure*}

Our final grade for each system ($\Gav$) is the average of the two grades resulting from the rapid and refined inspections outlined above. There are 26 candidates graded $\Gav\geq8$, detailed in \tref{tab:cand_G8NOT} with their CFIS cutouts and colours in \fref{fig:cand_G8NOT}. Before spectroscopic follow-up (\sref{sec:not}), we prioritise the candidates with detailed light modelling (\sref{sec:light}).

\begin{table*}[h]
  \caption{Lens candidates with $\Gav\geq8$ or with follow-up spectroscopy with ALFOSC on the NOT.}
  \centering
  \begin{tabular}{lrrccccccc}
  \hline
  \hline
  \multicolumn{1}{c}{Name} & \multicolumn{1}{c}{R.A. [$\deg$]} & \multicolumn{1}{c}{Dec. [$\deg$]} & $r_2-r_1$ & catalogue & $\Gav$ & $z_{\rm spec}$ & spectrograph & sep. [\arcsec] & outcome\\
  \hline
  UNIONS         J0112$+$3307 & $    18.198$ & $    33.132$ &     $0.35$ &         GD & $   9$ &      0.243 &     LAMOST &      1.036 &   INC\\
  UNIONS         J0800$+$4059 & $   120.011$ & $    40.991$ &     $1.57$ &         MQ & $   8$ &      1.622 &       SDSS &          - &  CONT\\
  UNIONS         J0805$+$3550 & $   121.273$ & $    35.847$ &     $0.39$ &         GD & $   9$ &      1.668 &       SDSS &      1.142 &   INC\\
  UNIONS         J0840$+$4631 & $   130.216$ & $    46.524$ &     $0.57$ &         MQ & $   8$ &      1.358 &       SDSS &          - &  CONT\\
  UNIONS         J0922$+$4028 & $   140.721$ & $    40.476$ &     $0.95$ &         MQ & $ 8.5$ &      1.280 &        NOT &      1.350 &   INC\\
  UNIONS         J0926$+$3059 & $   141.644$ & $    30.996$ &     $1.23$ &         MQ & $   8$ &      2.257 &       SDSS &      2.067 &   INC\\
  UNIONS         J0938$+$4216 & $   144.508$ & $    42.268$ &     $0.37$ &         MQ & $   9$ &      1.705 &        NOT &      0.797 &   INC\\
  UNIONS         J0943$+$3532 & $   145.862$ & $    35.544$ &     $0.39$ &      GD/MQ & $ 9.5$ &      1.210 &        NOT &      0.831 &  Lens\\
  UNIONS         J0958$+$4325 & $   149.656$ & $    43.432$ &     $0.89$ &         MQ & $   8$ &      0.813 &       SDSS &      0.985 &   INC\\
  UNIONS         J1006$+$3040 & $   151.576$ & $    30.677$ &     $0.82$ &      GD/MQ & $ 8.5$ &      1.345 &        NOT &      0.837 &   NIQ\\
  UNIONS         J1039$+$5126 & $   159.841$ & $    51.436$ &     $0.03$ &      GD/MQ & $ 6.5$ &      2.050 &        NOT &      1.693 &   NIQ\\
  UNIONS         J1111$+$3804 & $   167.777$ & $    38.073$ &     $0.51$ &         MQ & $  10$ &      3.020 &        NOT &      1.964 &  Lens\\
  UNIONS         J1136$+$4227 & $   174.086$ & $    42.458$ &     $2.05$ &         GD & $ 8.5$ &      1.800 &        NOT &      2.143 &   NIQ\\
  UNIONS         J1146$+$3254 & $   176.720$ & $    32.903$ &     $1.75$ &         MQ & $  10$ &      2.073 &        NOT &      2.441 &  Lens\\
  UNIONS         J1201$+$4013 & $   180.329$ & $    40.222$ &     $1.72$ &         MQ & $   9$ &      1.940 &        NOT &      1.609 &  Lens\\
  UNIONS         J1219$+$3453 & $   184.760$ & $    34.896$ &     $0.75$ &         MQ & $   9$ &      1.150 &        NOT &      4.150 &   NIQ\\
  UNIONS         J1223$+$3103 & $   185.839$ & $    31.054$ &     $1.98$ &         MQ & $ 8.5$ &      0.887 &        NOT &      1.011 &   INC\\
  UNIONS         J1325$+$3629 & $   201.339$ & $    36.487$ &     $1.46$ &      GD/MQ & $ 7.5$ &      1.210 &        NOT &      0.686 &   INC\\
  UNIONS         J1454$+$4146 & $   223.702$ & $    41.777$ &     $1.01$ &         MQ & $ 7.5$ &      1.670 &        NOT &      1.414 &    QP\\
  UNIONS         J1529$+$3841 & $   232.262$ & $    38.684$ &     $2.27$ &      GD/MQ & $   8$ &      2.012 &       SDSS &      0.918 &   INC\\
  UNIONS         J1537$+$3223 & $   234.254$ & $    32.391$ &     $0.05$ &      GD/MQ & $ 7.5$ &      1.825 &        NOT &      0.975 &   NIQ\\
  UNIONS         J1551$+$3423 & $   237.884$ & $    34.389$ &     $2.11$ &         MQ & $   8$ &      1.832 &       SDSS &      1.189 &   INC\\
  UNIONS         J1555$+$3118 & $   238.762$ & $    31.313$ &     $1.11$ &         GD & $   8$ &      3.360 &        NOT &      1.640 &  Lens\\
  UNIONS         J1555$+$3703 & $   238.997$ & $    37.061$ &     $1.69$ &         MQ & $ 8.5$ &      1.700 &        NOT &      0.926 &   INC\\
  UNIONS         J1555$+$3813 & $   238.983$ & $    38.226$ &     $0.66$ &         MQ & $ 8.5$ &      2.525 &        NOT &      0.749 &    QP\\
  UNIONS         J1559$+$3839 & $   239.816$ & $    38.653$ &     $0.75$ &         MQ & $   9$ &      1.334 &   SDSS-new &      0.885 & PLens\\
  UNIONS         J1633$+$4556 & $   248.287$ & $    45.942$ &     $1.70$ &         MQ & $   8$ &      1.412 &       SDSS &      0.825 &   INC\\
  UNIONS         J1655$+$3746 & $   253.774$ & $    37.777$ &     $0.80$ &         MQ & $   8$ &          - &          - &      1.045 &   INC\\
  UNIONS         J1739$+$4355 & $   264.848$ & $    43.919$ &     $0.14$ &         MQ & $   9$ &      2.515 &        NOT &      1.538 &   NIQ\\
  UNIONS         J2335$+$3201 & $   353.844$ & $    32.019$ &     $1.24$ &         MQ & $ 8.5$ &      0.903 &       SDSS &      0.620 &   INC\\
  \hline
  \end{tabular}
  \tablefoot{
  The flux ratio of brightest two blue point sources in $r$-band is denoted as $r_2-r_1$; see also \fref{fig:select}. The preselected quasar catalogues are denoted as `MQ' for MILLIQUAS and as `GD' for {\it Gaia} double detection; see \sref{subsec:catalog}. We list the average grade of the refined visual inspection as $\Gav$. The redshifts are measured through the spectra in the `spectrograph' column; see \ssref{sec:not} and \ref{sec:result}. The image separation is measured with light modelling. The outcomes indicate confirmed lenses as `Lens', probable lenses as `PLens', nearly identical quasars as `NIQ', quasar pairs as `QP', contaminants as `CONT', and inconclusive systems as `INC'.
  }
  \label{tab:cand_G8NOT}
\end{table*}

\section{Light modelling} 
\label{sec:light}

A lensed quasar system should contain a lensing galaxy surrounded by multiple lensed images.
In most cases, a lensing galaxy is fainter than a quasar and is barely to be seen in ground-based imaging. Although our classification method is able to highlight lensing galaxies by removing lensed quasar images (see \fref{fig:chitah} as an example), imperfect PSFs from CFIS data sometimes blend lens light and residuals, or residuals of double point sources can appear to mimic a lensing galaxy. We therefore attempt to optimally subtract the obvious point sources (or single point source and galaxy) in search of a lensing galaxy (or counter image) in the deep, sharp CFIS imaging for the most promising systems resulting from visual inspection.

A Moffat profile \citep{Moffat69} is used for the PSF (and hence also quasar images), and a S\'ersic profile \citep{Sersic63} for lensing galaxies. We manually decide the numbers of point sources and galaxies for each system, and fit the $u$ and $r$ bands simultaneously. The PSF parameters are fitted directly from the point sources in our candidates. Results from modelling the best candidates are shown in \fref{fig:cand_G8NOT} and \tref{tab:cand_mag}.

\section{Spectroscopic follow-up}
\label{sec:not}
Long-slit follow-up spectra were taken with the Alhambra Faint Object Spectrograph and Camera (ALFOSC) mounted on the $2.56$-m Nordic Optical Telescope (NOT) on the nights of 16 and 17 April 2021 with seeings of $\sim1.2\arcsec$ and $\sim0.75\arcsec$ respectively. To reliably identify a priori unknown quasar redshifts, we require multiple broad lines. We therefore used grism \#4 with a wide wavelength coverage of $3200~\AA$ -- $9600~\AA$ and dispersion of $3.3~\AA/\text{pixel}$.  Slit position angles were calculated from the best-fit image positions from the CFIS light modelling (\sref{sec:light}). The slit width was $1\arcsec$ and no atmospheric dispersion corrector was used to optimise throughput.

The 2D spectra were bias-subtracted and cosmic rays masked, and a model for the sky background based on the $5\arcsec$ either side of the object was subtracted. We extract separate spectra using a forward-modelling process, as described fully in \citet{LemonEtal21}. Briefly, we forward model the quasar images as Moffat profiles onto a pixelised grid, and derive the 1D spatial spread function after applying a slit of 1\arcsec\  in width. The Moffat parameters, image separation, and spatial variation along the slit are found in each wavelength bin using MCMC sampling. A second-order polynomial fit to these parameters then allows a model at each pixel in wavelength bins, followed by a least-squares fit to determine the flux. The residuals are inspected for goodness-of-fit and the spectra are binned into equal noise bins to facilitate comparison of spectra.

\begin{figure*}[h]
\centering
\includegraphics[scale=0.65]{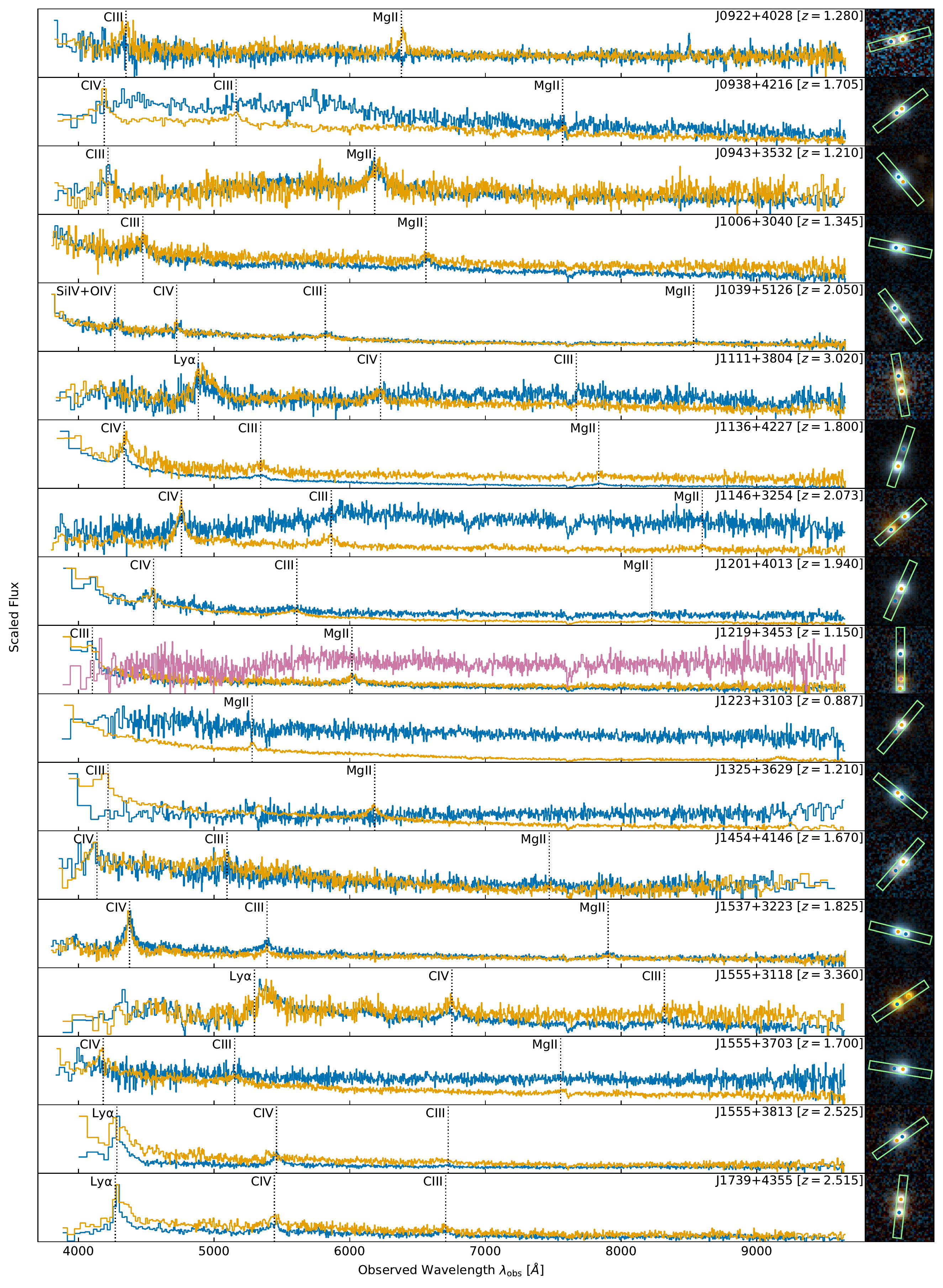}
\vskip -5pt
\caption{
Extracted 1D spectra of candidates observed with ALFOSC on the NOT. Redshifts are measured from the average of the peaks of broad emission lines. The CFIS colour image is shown to the right of each spectrum, with the slit position overlaid as a green box.
}
\label{fig:spec}
\end{figure*}

We prioritise our systems according to the visual inspection and the light modelling. In total we, obtained spectra for 18 systems, four of which have grades $\Gav<8$ due to our observing constraints. The spectra are shown in \fref{fig:spec}, and determined redshifts and classification are listed in \tref{tab:cand_G8NOT}.

\section{Result}
\label{sec:result}
Combining our spectroscopy and light modelling, the results for our best 30 lens candidates are summarised in \tref{tab:cand_G8NOT}. We note that redshifts are measured from either NOT or other existing spectrographs.

In this work, we confidently confirm five new lensed quasars: J0943+3532, J1111+3804, J1146+3254, J1201+4013, and J1555+3118. One further system, J1559+3839, is a highly probable lensed quasar; however, we did not obtain spatially resolved spectroscopy for this system. There are six systems that we designate as potential lensed quasars, as they show nearly identical quasar (NIQ) spectra in two components. Two systems show double quasar spectra at the same redshift, but different profiles in emission lines. We therefore categorise these latter two as quasar pairs. We also identify two contaminants in our best list, because we can only fit the light distribution with a single point source and a galaxy. Our results for the remainder of the sample are inconclusive.

We note that there are three systems with X-ray detections, namely J1325+3629, J1555+3813, and J1559+3839 (a probable lens), which can be used to study the inner structure of quasars \citep{Chartas00}. One of the NIQs and a possible lens, J1111+3804, which can be a possible lens, shows a radio detection. If this system is confirmed as a lens, we can exploit the radio flux ratio anomalies induced by substructures in lensing galaxies. Estimating the number of substructures allows us to discern diverse dark matter models \citep{HarveyEtal20}. The redshift range in our sample of lens candidates spans from 0.24 to 3.36 (from 1.21 to 3.36 for confirmed lenses), showing that our classification method is still able to detect lensed quasars at high redshift ($z\gtrsim2.7$), if there is no blue object located in the image cutout. Most of our candidates have small image separations, ranging from $0.6\arcsec$ to $4.2\arcsec$ (from $0.8\arcsec$ to $2.5\arcsec$ for confirmed lenses). The smallest separation is approximately equal to the seeing in CFIS-$r$, which can help to extend the sample size for lensing statistics. The lenses confirmed in this work are doubly lensed quasars, which can potentially serve as tools for the measurement of $H_0$ \citep{BirrerEtal19}. We do not find quadruply lensed quasars because they are missing in our quasar catalogues. The potential binary quasars found in this work, especially those with close separations ($\sim$kpc), can be used for the study of low-frequency gravitational waves from dual supermassive black holes \citep{GouldingEtal19,Casey-ClydeEtal21}, and those at high redshifts ($z\gtrsim2$) can even provide extra science cases with which to study the distant double quasars, such as dynamical evolution and successive mergers in the early Universe \citep{ShenEtal21}. We detailed a few interesting systems in the individual notes as follows.

\subsection{J0112+3307 (inconclusive)}
This system was given a high average grade of 9 (see Sect. 4), though was unfortunately not visible during our spectroscopic follow-up. We measure the redshift at $z=0.243$ using the LAMOST spectra \citep{LuoEtal15}, as shown in \fref{fig:spec_others}. Given the image separation and source redshift, we estimate the velocity dispersion $\sigma_v=250$~km/s (assuming lens redshift at $0.1$). Although a galaxy component can be detected from the light modelling (see \fref{fig:cand_G8NOT}), it seems too faint to achieve such high velocity dispersion \citep{BoltonEtal08}. We are therefore unable to classify this system as a lens. High-resolution spectroscopic or imaging data are needed to confirm its nature.

\subsection{J0943+3532 (lens)}
Given the small separation of $0.831\arcsec$ between the two components, the spectrum shows a single blended component, with emission lines of a quasar at $z=1.21$. {\it Gaia} detects two point-source components (with small excess noise and proper motion) and the CFIS image modelling requires an additional extended component between these two point sources (see \fref{fig:cand_G8NOT}). We therefore classify this system as a lens.

\subsection{J1006+3040 (NIQ)}
The two components have the same CIII and MgII lines. We noticed that H and K of CaII absorption and OII emission lines appear in the yellow spectrum at $z\sim0.3$. However, we are not able to see an extended component after fitting two point sources (see \fref{fig:cand_G8NOT}). This system requires higher resolution observations to confirm its lens feature.

\subsection{J1039+5126 (NIQ)}
SiIV+OIV, CIV, and CIII emission lines fit the same redshift at $z=2.05$. The slight deviation in CIV profiles could be attributed to microlensing. We therefore classify J1039+5126 as a NIQ.

\subsection{J1111+3804 (lens)}
Quasar source at high $z=3.02$ and lensing galaxy can be seen in light modelling. An existing SDSS spectrum centred on the lower quasar confirms the source redshift at $z=3.018$. The top component is redder in the spectrum (blue in \fref{fig:spec}), which suggests some light leaking from a galaxy. We can see the $4000~\AA$ break at H and K lines of CaII and G at $z\sim0.6$. We also note that there is a strong radio detection for this system from VLASS.

\subsection{J1146+3254 (lens)}
This new lens is an example of the power provided by the improved depth and seeing of CFIS in finding lensed quasars. A single point source and galaxy subtracted from Pan-STARRS or DECaLS\footnote{The DECaLS image of J1146+3254 can be seen in \url{https://www.legacysurvey.org/viewer?ra=176.720&dec=32.903&zoom=16}} images does not show any counter image, whereas in CFIS it is immediately apparent. Our resolved spectra detect the CIV line of the counter image blended with the lensing galaxy, which we measure to be at $z\sim0.3$.

\subsection{J1201+4013 (lens)}
This system was originally considered by the SDSS quasar lens search as a potential lens, SDSS J120118.92+401318.1, and it was selected as a morphological candidate with an SDSS QSO spectrum at $z=1.936$. \citet{InadaEtal10} ruled this out as a possible lens based on $I$-band imaging from Tek2k CCD at UH88 which was fit as a single point source+galaxy model. Thanks to the deeper and higher resolution spectra provided by {\it Gaia}, we selected this system as having two {\it Gaia} detections, and thus highly likely to have two 
point sources present in the system. Modelling of the CFIS data requires a counter image and a detection of the lensing galaxy as well (see \fref{fig:cand_G8NOT}). Resolved NOT spectra clearly show CIV and CIII emission from the fainter component, with a redder continuum, which is likely due to the lensing galaxy. We subtract the spectrum of A from B, and rescaled by the median flux ratio below $4800~\AA$. Excluding microlensing and differential reddening effects, this leaves the spectrum of the lensing galaxy, from which we can see absorption lines and a $4000~\AA$ break associated to $z\sim0.4$. 

\subsection{J1219+3453 (NIQ)}
Although two quasar images have the same redshift of $z=1.15$ and similar spectra, we are not able to confirm the nature of the central object (see the red line in \fref{fig:spec}); it is well fit by a PSF and is therefore likely a star. High-resolution imaging is required to determine the presence or lack of a lensing galaxy. We therefore classify this system as a NIQ instead of a lens.

\subsection{J1325+3629 (inconclusive)}
This system shows the X-ray detection in {\it Chandra} imaging. The faint component is likely to be a contaminant, because its spectrum does not have apparent emission lines. However, the very small separation makes it difficult to separate the 2D spectrum. We need high-resolution imaging or better resolved spectroscopy.

\subsection{J1454+4146 (quasar pair)}
Although two components have the same redshift, CIII and CIV profiles are not the same (see \fref{fig:spec}). We classify this system as a quasar pair.

\subsection{J1537+3223 (NIQ)}
Two components have very similar spectra at $z=1.825$. However, because of the small separation, we are not able to see a lensing galaxy. We classify this system as a NIQ. 

\subsection{J1555+3118 (lens)}
This system is a $u$-band dropout recovered by our selection at $z=3.360$ with a lensing galaxy detected. It was originally selected in the {\it Gaia} double-{\it WISE} cross-match catalogue and not in the MILLIQUAS sample because of its marginally bluer {\it WISE} colours of $W1-W2=0.67$ and $W2-W3=3.32$, typical of higher redshift quasars \citep{SecrestEtal15}. 

\subsection{J1555+3813 (quasar pair)}
Different Lyman alpha and CIV line profiles suggest that the two point sources are not from the same quasar source (see \fref{fig:spec}). In addition, we do not see an extended component between these two point sources (see \fref{fig:cand_G8NOT}). This system has a small image separation $0.749\arcsec$ (6.04~kpc at $z=2.525$). We notice that this system has X-ray detection in {\it ROSAT}.

\subsection{J1559+3839 (probable lens)}
Although we do not obtain the NOT follow-up spectra for this system, there is an SDSS spectrum with a QSO pipeline redshift of $z=4.37$. However, closer inspection of the spectrum reveals the main emission line is likely MgII with apparent CIII, NeV, and NeVI at $z=1.334$ (see \fref{fig:spec_others}). This is also the classified redshift of the improved classification of \citet{Hewett&Wild10}. The modelling residuals require a galaxy between the two point sources. We therefore classify it as a probable lensed quasar. The strong reddening of the background quasar can either be intrinsic, as in W2M J1042+1641 \citep{GlikmanEtal18} and MG 1131+0456 \citep{Stern&Walton20}, or due to the lensing galaxy. We also notice that this system has an X-ray detection in {\it ROSAT}.

\subsection{J1739+4355 (NIQ)}
Two quasars have similar spectra with a small image separation $1.538\arcsec$ (12.4~kpc at $z=2.525$). We notice that there is a big difference in Ly$\alpha$ or in the NV line ($1240.22~\AA$). The deviation of Ly$\alpha$ could be attributed to microlensing, so we still classify this system as a NIQ/possible lens.

\section{Conclusion} 
\label{sec:conclusion}

In this work, we present new lens candidates in CFIS DR2 (a component of UNIONS), covering $2\,500\deg^2$ with depths of $u=23.6$~mag and $r=24.1$~mag. We confirm the lens features based on light modelling and spectroscopic follow-up. We draw the following conclusions:
\begin{itemize}
    \item We begin from the catalogues with MILLIQUAS and {\it Gaia} pairs, and then extract the image cutouts in CFIS DR2 with $u$ and $r$ bands. 
    The total number of objects is 256\,314.
    \item We designed an efficient classification method: searching for a system as a lens candidate that contains multiple blue point sources, from 2 to 5  in number. This method is able to recover 15 of the 18 known lenses.
    \item We classified 10\,914 candidates with the classification method, and further removed $20\%$ of candidates with the cut of $r$-mag difference with $|r_1-r_2|<2.5$~mag.
    \item We visually inspected 7\,815 candidates, finding 239 promising ones which we then graded from 0 (non-lens) to 10 (lens) according to criteria we outline in Sect. 4.
    \item We carried out a follow-up analysis of 30 high-grade candidates with light modelling and 18 of them are observed with ALFOSC on the NOT.
    \item We confirm five new lenses (J0943+3532, J1111+3804, J1146+3254, J1201+4013, and J1555+3118) and one probable lens (J1559+3839). Six systems are nearly identical quasars, which are potential lensed quasars. Two systems are classified as quasar pairs, showing double quasar spectra but different emission line profiles. We also identified two contaminants during the light modelling.
\end{itemize}
CFIS imaging data allow us to see more lenses of small separation (e.g. the smallest one is $0.831\arcsec$ in this work), thanks to its high resolution in $r$ band. The full CFIS sample will continue to be analysed as part of a future exploration. Upcoming data releases will contain deep and high-resolution imaging in the $u$, $g$, $r$, $i$ and $z$ bands from all the components of the UNIONS collaboration. The $g$ and $z$ bands obtained with Subaru/HSC will also allow us to make the most of the benefits of high angular resolution. The success of this work demonstrates that the classification method can be applied to the higher resolution imaging data in future, such as Euclid and the Vera C. Rubin Observatory's Legacy Survey of Space and Time (LSST), and will therefore be exceptionally useful for finding the lenses with the narrowest separation on the sky.


\section*{Acknowledgements}
This work is supported by the Swiss National Science Foundation (SNSF) and by the European Research Council (ERC) under the European Union's Horizon 2020 research and innovation program (COSMICLENS: grant agreement No 787886).
Part of this work benefited from the Canadian Advanced Network for Astronomical Research (CANFAR) and Compute Canada facilities.
R.~G. thanks IoA and the Churchill College in Cambridge for their hospitality and acknowledges local support from the French government.
G.~V. has received funding from the European Union's Horizon 2020 research and innovation programme under the Marie Sk{\l}odovska-Curie grant agreement No. 897124.
This work is based on data obtained as part of the Canada-France Imaging Survey, a CFHT large program of the National Research Council of Canada and the French Centre National de la Recherche Scientifique. Based on observations obtained with MegaPrime/MegaCam, a joint project of CFHT and CEA Saclay, at the Canada-France-Hawaii Telescope (CFHT) which is operated by the National Research Council (NRC) of Canada, the Institut National des Science de l'Univers (INSU) of the Centre National de la Recherche Scientifique (CNRS) of France, and the University of Hawaii. This research used the facilities of the Canadian Astronomy Data Centre operated by the National Research Council of Canada with the support of the Canadian Space Agency. This research is based in part on data collected at Subaru Telescope, which is operated by the National Astronomical Observatory of Japan. We are honored and grateful for the opportunity of observing the Universe from Maunakea, which has the cultural, historical and natural significance in Hawaii. Pan-STARRS is a project of the Institute for Astronomy of the University of Hawaii, and is supported by the NASA SSO Near Earth Observation Program under grants 80NSSC18K0971, NNX14AM74G, NNX12AR65G, NNX13AQ47G, NNX08AR22G, YORPD20\_2-0014 and by the State of Hawaii.


\bibliographystyle{aa}
\bibliography{reference}

\begin{thebibliography}{62}
\expandafter\ifx\csname natexlab\endcsname\relax\def\natexlab#1{#1}\fi

\bibitem[{{Aihara} {et~al.}(2018){Aihara}, {Armstrong}, {Bickerton}, {Bosch},
  {Coupon}, {Furusawa}, {Hayashi}, {Ikeda}, {Kamata}, {Karoji}, {Kawanomoto},
  {Koike}, {Komiyama}, {Lang}, {Lupton}, {Mineo}, {Miyatake}, {Miyazaki},
  {Morokuma}, {Obuchi}, {Oishi}, {Okura}, {Price}, {Takata}, {Tanaka},
  {Tanaka}, {Tanaka}, {Uchida}, {Uraguchi}, {Utsumi}, {Wang}, {Yamada},
  {Yamanoi}, {Yasuda}, {Arimoto}, {Chiba}, {Finet}, {Fujimori}, {Fujimoto},
  {Furusawa}, {Goto}, {Goulding}, {Gunn}, {Harikane}, {Hattori}, {Hayashi},
  {He{\l}miniak}, {Higuchi}, {Hikage}, {Ho}, {Hsieh}, {Huang}, {Huang},
  {Imanishi}, {Iwata}, {Jaelani}, {Jian}, {Kashikawa}, {Katayama}, {Kojima},
  {Konno}, {Koshida}, {Kusakabe}, {Leauthaud}, {Lee}, {Lin}, {Lin},
  {Mandelbaum}, {Matsuoka}, {Medezinski}, {Miyama}, {Momose}, {More}, {More},
  {Mukae}, {Murata}, {Murayama}, {Nagao}, {Nakata}, {Niida}, {Niikura},
  {Nishizawa}, {Oguri}, {Okabe}, {Ono}, {Onodera}, {Onoue}, {Ouchi}, {Pyo},
  {Shibuya}, {Shimasaku}, {Simet}, {Speagle}, {Spergel}, {Strauss}, {Sugahara},
  {Sugiyama}, {Suto}, {Suzuki}, {Tait}, {Takada}, {Terai}, {Toba}, {Turner},
  {Uchiyama}, {Umetsu}, {Urata}, {Usuda}, {Yeh}, \& {Yuma}}]{AiharaEtal18}
{Aihara}, H., {Armstrong}, R., {Bickerton}, S., {et~al.} 2018, \pasj, 70, S8

\bibitem[{{Bertin}(2013)}]{Bertin13}
{Bertin}, E. 2013, {PSFEx: Point Spread Function Extractor}

\bibitem[{{Birrer} {et~al.}(2019){Birrer}, {Treu}, {Rusu}, {Bonvin},
  {Fassnacht}, {Chan}, {Agnello}, {Shajib}, {Chen}, {Auger}, {Courbin},
  {Hilbert}, {Sluse}, {Suyu}, {Wong}, {Marshall}, {Lemaux}, \&
  {Meylan}}]{BirrerEtal19}
{Birrer}, S., {Treu}, T., {Rusu}, C.~E., {et~al.} 2019, \mnras, 484, 4726

\bibitem[{{Bolton} {et~al.}(2008){Bolton}, {Burles}, {Koopmans}, {Treu},
  {Gavazzi}, {Moustakas}, {Wayth}, \& {Schlegel}}]{BoltonEtal08}
{Bolton}, A.~S., {Burles}, S., {Koopmans}, L. V.~E., {et~al.} 2008, \apj, 682,
  964

\bibitem[{{Cantale} {et~al.}(2016){Cantale}, {Courbin}, {Tewes}, {Jablonka}, \&
  {Meylan}}]{CantaleEtal16}
{Cantale}, N., {Courbin}, F., {Tewes}, M., {Jablonka}, P., \& {Meylan}, G.
  2016, \aap, 589, A81

\bibitem[{{Casey-Clyde} {et~al.}(2021){Casey-Clyde}, {Mingarelli}, {Greene},
  {Pardo}, {Na{\~n}ez}, \& {Goulding}}]{Casey-ClydeEtal21}
{Casey-Clyde}, J.~A., {Mingarelli}, C. M.~F., {Greene}, J.~E., {et~al.} 2021,
  arXiv e-prints, arXiv:2107.11390

\bibitem[{{Chan} {et~al.}(2015){Chan}, {Suyu}, {Chiueh}, {More}, {Marshall},
  {Coupon}, {Oguri}, \& {Price}}]{ChanEtal15}
{Chan}, J.~H.~H., {Suyu}, S.~H., {Chiueh}, T., {et~al.} 2015, \apj, 807, 138

\bibitem[{{Chan} {et~al.}(2020){Chan}, {Suyu}, {Sonnenfeld}, {Jaelani}, {More},
  {Yonehara}, {Kubota}, {Coupon}, {Lee}, {Oguri}, {Rusu}, \&
  {Wong}}]{ChanEtal20}
{Chan}, J. H.~H., {Suyu}, S.~H., {Sonnenfeld}, A., {et~al.} 2020, \aap, 636,
  A87

\bibitem[{{Chartas}(2000)}]{Chartas00}
{Chartas}, G. 2000, \apj, 531, 81

\bibitem[{{Chen} {et~al.}(2019){Chen}, {Fassnacht}, {Suyu}, {Rusu}, {Chan},
  {Wong}, {Auger}, {Hilbert}, {Bonvin}, {Birrer}, {Millon}, {Koopmans},
  {Lagattuta}, {McKean}, {Vegetti}, {Courbin}, {Ding}, {Halkola}, {Jee},
  {Shajib}, {Sluse}, {Sonnenfeld}, \& {Treu}}]{ChenEtal19}
{Chen}, G. C.~F., {Fassnacht}, C.~D., {Suyu}, S.~H., {et~al.} 2019, \mnras,
  490, 1743

\bibitem[{{Dalal} \& {Kochanek}(2002)}]{Dalal&Kochanek02}
{Dalal}, N. \& {Kochanek}, C.~S. 2002, \apj, 572, 25

\bibitem[{{Dey} {et~al.}(2019){Dey}, {Schlegel}, {Lang}, {Blum}, {Burleigh},
  {Fan}, {Findlay}, {Finkbeiner}, {Herrera}, {Juneau}, {Landriau}, {Levi},
  {McGreer}, {Meisner}, {Myers}, {Moustakas}, {Nugent}, {Patej}, {Schlafly},
  {Walker}, {Valdes}, {Weaver}, {Y{\`e}che}, {Zou}, {Zhou}, {Abareshi},
  {Abbott}, {Abolfathi}, {Aguilera}, {Alam}, {Allen}, {Alvarez}, {Annis},
  {Ansarinejad}, {Aubert}, {Beechert}, {Bell}, {BenZvi}, {Beutler}, {Bielby},
  {Bolton}, {Brice{\~n}o}, {Buckley-Geer}, {Butler}, {Calamida}, {Carlberg},
  {Carter}, {Casas}, {Castander}, {Choi}, {Comparat}, {Cukanovaite}, {Delubac},
  {DeVries}, {Dey}, {Dhungana}, {Dickinson}, {Ding}, {Donaldson}, {Duan},
  {Duckworth}, {Eftekharzadeh}, {Eisenstein}, {Etourneau}, {Fagrelius},
  {Farihi}, {Fitzpatrick}, {Font-Ribera}, {Fulmer}, {G{\"a}nsicke},
  {Gaztanaga}, {George}, {Gerdes}, {Gontcho}, {Gorgoni}, {Green}, {Guy},
  {Harmer}, {Hernandez}, {Honscheid}, {Huang}, {James}, {Jannuzi}, {Jiang},
  {Joyce}, {Karcher}, {Karkar}, {Kehoe}, {Kneib}, {Kueter-Young}, {Lan},
  {Lauer}, {Le Guillou}, {Le Van Suu}, {Lee}, {Lesser}, {Perreault Levasseur},
  {Li}, {Mann}, {Marshall}, {Mart{\'\i}nez-V{\'a}zquez}, {Martini}, {du Mas des
  Bourboux}, {McManus}, {Meier}, {M{\'e}nard}, {Metcalfe},
  {Mu{\~n}oz-Guti{\'e}rrez}, {Najita}, {Napier}, {Narayan}, {Newman}, {Nie},
  {Nord}, {Norman}, {Olsen}, {Paat}, {Palanque-Delabrouille}, {Peng},
  {Poppett}, {Poremba}, {Prakash}, {Rabinowitz}, {Raichoor}, {Rezaie},
  {Robertson}, {Roe}, {Ross}, {Ross}, {Rudnick}, {Safonova}, {Saha},
  {S{\'a}nchez}, {Savary}, {Schweiker}, {Scott}, {Seo}, {Shan}, {Silva},
  {Slepian}, {Soto}, {Sprayberry}, {Staten}, {Stillman}, {Stupak}, {Summers},
  {Sien Tie}, {Tirado}, {Vargas-Maga{\~n}a}, {Vivas}, {Wechsler}, {Williams},
  {Yang}, {Yang}, {Yapici}, {Zaritsky}, {Zenteno}, {Zhang}, {Zhang}, {Zhou}, \&
  {Zhou}}]{DeyEtal19}
{Dey}, A., {Schlegel}, D.~J., {Lang}, D., {et~al.} 2019, \aj, 157, 168

\bibitem[{{Ducourant} {et~al.}(2018{\natexlab{a}}){Ducourant}, {Delchambre},
  {Finet}, {Galluccio}, {Krone-Martins}, {Le Campion}, {Mignard}, {Slezak},
  {Surdej}, {Teixeira}, \& {Wertz}}]{DucourantEtal18a}
{Ducourant}, C., {Delchambre}, L., {Finet}, F., {et~al.} 2018{\natexlab{a}}, in
  IAU Symposium, Vol. 330, Astrometry and Astrophysics in the Gaia Sky, ed.
  A.~{Recio-Blanco}, P.~{de Laverny}, A.~G.~A. {Brown}, \& T.~{Prusti}, 59--62

\bibitem[{{Ducourant} {et~al.}(2018{\natexlab{b}}){Ducourant}, {Wertz},
  {Krone-Martins}, {Teixeira}, {Le Campion}, {Galluccio}, {Kl{\"u}ter},
  {Delchambre}, {Surdej}, {Mignard}, {Wambsganss}, {Bastian}, {Graham},
  {Djorgovski}, \& {Slezak}}]{DucourantEtal18b}
{Ducourant}, C., {Wertz}, O., {Krone-Martins}, A., {et~al.} 2018{\natexlab{b}},
  \aap, 618, A56

\bibitem[{{Fantin} {et~al.}(2019){Fantin}, {C{\^o}t{\'e}}, {McConnachie},
  {Bergeron}, {Cuillandre}, {Gwyn}, {Ibata}, {Thomas}, {Carlberg}, {Fabbro},
  {Haywood}, {Lan{\c{c}}on}, {Lewis}, {Malhan}, {Martin}, {Navarro}, {Scott},
  \& {Starkenburg}}]{FantinEtal19}
{Fantin}, N.~J., {C{\^o}t{\'e}}, P., {McConnachie}, A.~W., {et~al.} 2019, \apj,
  887, 148

\bibitem[{{Finlator} {et~al.}(2000){Finlator}, {Ivezi{\'c}}, {Fan}, {Strauss},
  {Knapp}, {Lupton}, {Gunn}, {Rockosi}, {Anderson}, {Csabai}, {Hennessy},
  {Hindsley}, {McKay}, {Nichol}, {Schneider}, {Smith}, {York}, \& {SDSS
  Collaboration}}]{FinlatorEtal20}
{Finlator}, K., {Ivezi{\'c}}, {\v{Z}}., {Fan}, X., {et~al.} 2000, \aj, 120,
  2615

\bibitem[{{Flesch}(2019)}]{Flesch19}
{Flesch}, E.~W. 2019, arXiv e-prints, arXiv:1912.05614

\bibitem[{{Gaia Collaboration} {et~al.}(2018){Gaia Collaboration}, {Brown},
  {Vallenari}, {Prusti}, {de Bruijne}, {Babusiaux}, {Bailer-Jones}, {Biermann},
  {Evans}, {Eyer}, {Jansen}, {Jordi}, {Klioner}, {Lammers}, {Lindegren},
  {Luri}, {Mignard}, {Panem}, {Pourbaix}, {Randich}, {Sartoretti}, {Siddiqui},
  {Soubiran}, {van Leeuwen}, {Walton}, {Arenou}, {Bastian}, {Cropper},
  {Drimmel}, {Katz}, {Lattanzi}, {Bakker}, {Cacciari}, {Casta{\~n}eda},
  {Chaoul}, {Cheek}, {De Angeli}, {Fabricius}, {Guerra}, {Holl}, {Masana},
  {Messineo}, {Mowlavi}, {Nienartowicz}, {Panuzzo}, {Portell}, {Riello},
  {Seabroke}, {Tanga}, {Th{\'e}venin}, {Gracia-Abril}, {Comoretto},
  {Garcia-Reinaldos}, {Teyssier}, {Altmann}, {Andrae}, {Audard},
  {Bellas-Velidis}, {Benson}, {Berthier}, {Blomme}, {Burgess}, {Busso},
  {Carry}, {Cellino}, {Clementini}, {Clotet}, {Creevey}, {Davidson}, {De
  Ridder}, {Delchambre}, {Dell'Oro}, {Ducourant},
  {Fern{\'a}ndez-Hern{\'a}ndez}, {Fouesneau}, {Fr{\'e}mat}, {Galluccio},
  {Garc{\'\i}a-Torres}, {Gonz{\'a}lez-N{\'u}{\~n}ez}, {Gonz{\'a}lez-Vidal},
  {Gosset}, {Guy}, {Halbwachs}, {Hambly}, {Harrison}, {Hern{\'a}ndez},
  {Hestroffer}, {Hodgkin}, {Hutton}, {Jasniewicz}, {Jean-Antoine-Piccolo},
  {Jordan}, {Korn}, {Krone-Martins}, {Lanzafame}, {Lebzelter}, {L{\"o}ffler},
  {Manteiga}, {Marrese}, {Mart{\'\i}n-Fleitas}, {Moitinho}, {Mora}, {Muinonen},
  {Osinde}, {Pancino}, {Pauwels}, {Petit}, {Recio-Blanco}, {Richards},
  {Rimoldini}, {Robin}, {Sarro}, {Siopis}, {Smith}, {Sozzetti}, {S{\"u}veges},
  {Torra}, {van Reeven}, {Abbas}, {Abreu Aramburu}, {Accart}, {Aerts},
  {Altavilla}, {{\'A}lvarez}, {Alvarez}, {Alves}, {Anderson}, {Andrei},
  {Anglada Varela}, {Antiche}, {Antoja}, {Arcay}, {Astraatmadja}, {Bach},
  {Baker}, {Balaguer-N{\'u}{\~n}ez}, {Balm}, {Barache}, {Barata}, {Barbato},
  {Barblan}, {Barklem}, {Barrado}, {Barros}, {Barstow}, {Bartholom{\'e}
  Mu{\~n}oz}, {Bassilana}, {Becciani}, {Bellazzini}, {Berihuete}, {Bertone},
  {Bianchi}, {Bienaym{\'e}}, {Blanco-Cuaresma}, {Boch}, {Boeche}, {Bombrun},
  {Borrachero}, {Bossini}, {Bouquillon}, {Bourda}, {Bragaglia}, {Bramante},
  {Breddels}, {Bressan}, {Brouillet}, {Br{\"u}semeister}, {Brugaletta},
  {Bucciarelli}, {Burlacu}, {Busonero}, {Butkevich}, {Buzzi}, {Caffau},
  {Cancelliere}, {Cannizzaro}, {Cantat-Gaudin}, {Carballo}, {Carlucci},
  {Carrasco}, {Casamiquela}, {Castellani}, {Castro-Ginard}, {Charlot},
  {Chemin}, {Chiavassa}, {Cocozza}, {Costigan}, {Cowell}, {Crifo}, {Crosta},
  {Crowley}, {Cuypers}, {Dafonte}, {Damerdji}, {Dapergolas}, {David}, {David},
  {de Laverny}, {De Luise}, {De March}, {de Martino}, {de Souza}, {de Torres},
  {Debosscher}, {del Pozo}, {Delbo}, {Delgado}, {Delgado}, {Di Matteo},
  {Diakite}, {Diener}, {Distefano}, {Dolding}, {Drazinos}, {Dur{\'a}n},
  {Edvardsson}, {Enke}, {Eriksson}, {Esquej}, {Eynard Bontemps}, {Fabre},
  {Fabrizio}, {Faigler}, {Falc{\~a}o}, {Farr{\`a}s Casas}, {Federici},
  {Fedorets}, {Fernique}, {Figueras}, {Filippi}, {Findeisen}, {Fonti},
  {Fraile}, {Fraser}, {Fr{\'e}zouls}, {Gai}, {Galleti}, {Garabato},
  {Garc{\'\i}a-Sedano}, {Garofalo}, {Garralda}, {Gavel}, {Gavras}, {Gerssen},
  {Geyer}, {Giacobbe}, {Gilmore}, {Girona}, {Giuffrida}, {Glass}, {Gomes},
  {Granvik}, {Gueguen}, {Guerrier}, {Guiraud}, {Guti{\'e}rrez-S{\'a}nchez},
  {Haigron}, {Hatzidimitriou}, {Hauser}, {Haywood}, {Heiter}, {Helmi}, {Heu},
  {Hilger}, {Hobbs}, {Hofmann}, {Holland}, {Huckle}, {Hypki}, {Icardi},
  {Jan{\ss}en}, {Jevardat de Fombelle}, {Jonker}, {Juh{\'a}sz}, {Julbe},
  {Karampelas}, {Kewley}, {Klar}, {Kochoska}, {Kohley}, {Kolenberg},
  {Kontizas}, {Kontizas}, {Koposov}, {Kordopatis}, {Kostrzewa-Rutkowska},
  {Koubsky}, {Lambert}, {Lanza}, {Lasne}, {Lavigne}, {Le Fustec}, {Le
  Poncin-Lafitte}, {Lebreton}, {Leccia}, {Leclerc}, {Lecoeur-Taibi},
  {Lenhardt}, {Leroux}, {Liao}, {Licata}, {Lindstr{\o}m}, {Lister}, {Livanou},
  {Lobel}, {L{\'o}pez}, {Managau}, {Mann}, {Mantelet}, {Marchal}, {Marchant},
  {Marconi}, {Marinoni}, {Marschalk{\'o}}, {Marshall}, {Martino}, {Marton},
  {Mary}, {Massari}, {Matijevi{\v{c}}}, {Mazeh}, {McMillan}, {Messina},
  {Michalik}, {Millar}, {Molina}, {Molinaro}, {Moln{\'a}r}, {Montegriffo},
  {Mor}, {Morbidelli}, {Morel}, {Morris}, {Mulone}, {Muraveva}, {Musella},
  {Nelemans}, {Nicastro}, {Noval}, {O'Mullane}, {Ord{\'e}novic},
  {Ord{\'o}{\~n}ez-Blanco}, {Osborne}, {Pagani}, {Pagano}, {Pailler},
  {Palacin}, {Palaversa}, {Panahi}, {Pawlak}, {Piersimoni}, {Pineau}, {Plachy},
  {Plum}, {Poggio}, {Poujoulet}, {Pr{\v{s}}a}, {Pulone}, {Racero}, {Ragaini},
  {Rambaux}, {Ramos-Lerate}, {Regibo}, {Reyl{\'e}}, {Riclet}, {Ripepi}, {Riva},
  {Rivard}, {Rixon}, {Roegiers}, {Roelens}, {Romero-G{\'o}mez}, {Rowell},
  {Royer}, {Ruiz-Dern}, {Sadowski}, {Sagrist{\`a} Sell{\'e}s}, {Sahlmann},
  {Salgado}, {Salguero}, {Sanna}, {Santana-Ros}, {Sarasso}, {Savietto},
  {Schultheis}, {Sciacca}, {Segol}, {Segovia}, {S{\'e}gransan}, {Shih},
  {Siltala}, {Silva}, {Smart}, {Smith}, {Solano}, {Solitro}, {Sordo}, {Soria
  Nieto}, {Souchay}, {Spagna}, {Spoto}, {Stampa}, {Steele},
  {Steidelm{\"u}ller}, {Stephenson}, {Stoev}, {Suess}, {Surdej}, {Szabados},
  {Szegedi-Elek}, {Tapiador}, {Taris}, {Tauran}, {Taylor}, {Teixeira},
  {Terrett}, {Teyssandier}, {Thuillot}, {Titarenko}, {Torra Clotet}, {Turon},
  {Ulla}, {Utrilla}, {Uzzi}, {Vaillant}, {Valentini}, {Valette}, {van Elteren},
  {Van Hemelryck}, {van Leeuwen}, {Vaschetto}, {Vecchiato}, {Veljanoski},
  {Viala}, {Vicente}, {Vogt}, {von Essen}, {Voss}, {Votruba}, {Voutsinas},
  {Walmsley}, {Weiler}, {Wertz}, {Wevers}, {Wyrzykowski}, {Yoldas},
  {{\v{Z}}erjal}, {Ziaeepour}, {Zorec}, {Zschocke}, {Zucker}, {Zurbach}, \&
  {Zwitter}}]{gaiadr2}
{Gaia Collaboration}, {Brown}, A.~G.~A., {Vallenari}, A., {et~al.} 2018, \aap,
  616, A1

\bibitem[{{Gilman} {et~al.}(2019){Gilman}, {Birrer}, {Treu}, {Nierenberg}, \&
  {Benson}}]{GilmanEtal19}
{Gilman}, D., {Birrer}, S., {Treu}, T., {Nierenberg}, A., \& {Benson}, A. 2019,
  \mnras, 1618

\bibitem[{{Glikman} {et~al.}(2018){Glikman}, {Rusu}, {Djorgovski}, {Graham},
  {Stern}, {Urrutia}, {Lacy}, \& {O'Meara}}]{GlikmanEtal18}
{Glikman}, E., {Rusu}, C.~E., {Djorgovski}, S.~G., {et~al.} 2018, arXiv
  e-prints, arXiv:1807.05434

\bibitem[{{Goulding} {et~al.}(2019){Goulding}, {Pardo}, {Greene}, {Mingarelli},
  {Nyland}, \& {Strauss}}]{GouldingEtal19}
{Goulding}, A.~D., {Pardo}, K., {Greene}, J.~E., {et~al.} 2019, \apjl, 879, L21

\bibitem[{{Guinot} {et~al.}(2021){Guinot}, {Kilbinger}, {Farrens}, {Peel},
  {Pujol}, {Schmitz}, {Starck}, {Erben}, {Gavazzi}, {Hudson}, {Hildebrandt},
  {Liaudat}, {Miller}, {Spitzer}, {Van Waerbeke}, {Cuillandre}, {Fabbro},
  {McConnachie}, \& {Mellier}}]{GuinotEtal21}
{Guinot}, A., {Kilbinger}, M., {Farrens}, S., {et~al.} 2021, \aap, submitted

\bibitem[{{Gwyn}(2019)}]{Gwyn09}
{Gwyn}, S. 2019, in Astronomical Society of the Pacific Conference Series, Vol.
  523, Astronomical Data Analysis Software and Systems XXVII, ed. P.~J.
  {Teuben}, M.~W. {Pound}, B.~A. {Thomas}, \& E.~M. {Warner}, 649

\bibitem[{{Gwyn}(2008)}]{Gwyn08}
{Gwyn}, S. D.~J. 2008, \pasp, 120, 212

\bibitem[{{Harvey} {et~al.}(2020){Harvey}, {Valkenburg}, {Tamone}, {Boyarsky},
  {Courbin}, \& {Lovell}}]{HarveyEtal20}
{Harvey}, D., {Valkenburg}, W., {Tamone}, A., {et~al.} 2020, \mnras, 491, 4247

\bibitem[{{Hewett} \& {Wild}(2010)}]{Hewett&Wild10}
{Hewett}, P.~C. \& {Wild}, V. 2010, \mnras, 405, 2302

\bibitem[{{Ibata} {et~al.}(2017){Ibata}, {McConnachie}, {Cuillandre}, {Fantin},
  {Haywood}, {Martin}, {Bergeron}, {Beckmann}, {Bernard}, {Bonifacio},
  {Caffau}, {Carlberg}, {C{\^o}t{\'e}}, {Cabanac}, {Chapman}, {Duc}, {Durret},
  {Famaey}, {Fabbro}, {Gwyn}, {Hammer}, {Hill}, {Hudson}, {Lan{\c{c}}on},
  {Lewis}, {Malhan}, {di Matteo}, {McCracken}, {Mei}, {Mellier}, {Navarro},
  {Pires}, {Pritchet}, {Reyl{\'e}}, {Richer}, {Robin}, {S{\'a}nchez-Janssen},
  {Sawicki}, {Scott}, {Scottez}, {Spekkens}, {Starkenburg}, {Thomas}, \&
  {Venn}}]{IbataEtal17}
{Ibata}, R.~A., {McConnachie}, A., {Cuillandre}, J.-C., {et~al.} 2017, \apj,
  848, 128

\bibitem[{{Inada} {et~al.}(2008){Inada}, {Oguri}, {Becker}, {Shin}, {Richards},
  {Hennawi}, {White}, {Pindor}, {Strauss}, \& {Kochanek}}]{InadaEtal08}
{Inada}, N., {Oguri}, M., {Becker}, R.~H., {et~al.} 2008, \aj, 135, 496

\bibitem[{{Inada} {et~al.}(2010){Inada}, {Oguri}, {Shin}, {Kayo}, {Strauss},
  {Hennawi}, {Morokuma}, {Becker}, {White}, \& {Kochanek}}]{InadaEtal10}
{Inada}, N., {Oguri}, M., {Shin}, M.-S., {et~al.} 2010, \aj, 140, 403

\bibitem[{{Inada} {et~al.}(2012){Inada}, {Oguri}, {Shin}, {Kayo}, {Strauss},
  {Morokuma}, {Rusu}, {Fukugita}, {Kochanek}, \& {Richards}}]{InadaEtal12}
{Inada}, N., {Oguri}, M., {Shin}, M.-S., {et~al.} 2012, \aj, 143, 119

\bibitem[{{Jackson} {et~al.}(2012){Jackson}, {Rampadarath}, {Ofek}, {Oguri}, \&
  {Shin}}]{JacksonEtal12}
{Jackson}, N., {Rampadarath}, H., {Ofek}, E.~O., {Oguri}, M., \& {Shin}, M.-S.
  2012, \mnras, 419, 2014

\bibitem[{{Jaelani} {et~al.}(2020){Jaelani}, {More}, {Oguri}, {Sonnenfeld},
  {Suyu}, {Rusu}, {Wong}, {Chan}, {Kayo}, {Lee}, {Chao}, {Coupon}, {Inoue}, \&
  {Futamase}}]{JaelaniEtal20}
{Jaelani}, A.~T., {More}, A., {Oguri}, M., {et~al.} 2020, \mnras, 495, 1291

\bibitem[{{Jaelani} {et~al.}(2021){Jaelani}, {Rusu}, {Kayo}, {More},
  {Sonnenfeld}, {Silverman}, {Schramm}, {Anguita}, {Inada}, {Kondo},
  {Schechter}, {Lee}, {Oguri}, {Chan}, {Wong}, \& {Inoue}}]{JaelaniEtal21}
{Jaelani}, A.~T., {Rusu}, C.~E., {Kayo}, I., {et~al.} 2021, \mnras, 502, 1487

\bibitem[{{Krone-Martins} {et~al.}(2018){Krone-Martins}, {Delchambre}, {Wertz},
  {Ducourant}, {Mignard}, {Teixeira}, {Kl{\"u}ter}, {Le Campion}, {Galluccio},
  {Surdej}, {Bastian}, {Wambsganss}, {Graham}, {Djorgovski}, \&
  {Slezak}}]{Krone-MartinsEtal18}
{Krone-Martins}, A., {Delchambre}, L., {Wertz}, O., {et~al.} 2018, \aap, 616,
  L11

\bibitem[{{Lemon} {et~al.}(2021){Lemon}, {Millon}, {Sluse}, {Courbin}, {Auger},
  {Chan}, {Paic}, \& {Agnello}}]{LemonEtal21}
{Lemon}, C., {Millon}, M., {Sluse}, D., {et~al.} 2021, arXiv e-prints,
  arXiv:2109.01144

\bibitem[{{Lemon} {et~al.}(2019){Lemon}, {Auger}, \& {McMahon}}]{LemonEtal19}
{Lemon}, C.~A., {Auger}, M.~W., \& {McMahon}, R.~G. 2019, \mnras, 483, 4242

\bibitem[{{Lemon} {et~al.}(2018){Lemon}, {Auger}, {McMahon}, \&
  {Ostrovski}}]{LemonEtal18}
{Lemon}, C.~A., {Auger}, M.~W., {McMahon}, R.~G., \& {Ostrovski}, F. 2018,
  \mnras, 479, 5060

\bibitem[{{Luo} {et~al.}(2015){Luo}, {Zhao}, {Zhao}, {Deng}, {Liu}, {Jing},
  {Wang}, {Zhang}, {Shi}, {Cui}, {Chu}, {Li}, {Bai}, {Wu}, {Cai}, {Cao}, {Cao},
  {Carlin}, {Chen}, {Chen}, {Chen}, {Chen}, {Chen}, {Chen}, {Chen},
  {Christlieb}, {Chu}, {Cui}, {Dong}, {Du}, {Fan}, {Feng}, {Fu}, {Gao}, {Gong},
  {Gu}, {Guo}, {Han}, {He}, {Hou}, {Hou}, {Hou}, {Hu}, {Hu}, {Hu}, {Huo},
  {Jia}, {Jiang}, {Jiang}, {Jiang}, {Jin}, {Kong}, {Kong}, {Lei}, {Li}, {Li},
  {Li}, {Li}, {Li}, {Li}, {Li}, {Li}, {Li}, {Li}, {Li}, {Li}, {Liang}, {Lin},
  {Liu}, {Liu}, {Liu}, {Liu}, {Lu}, {Luo}, {Mao}, {Newberg}, {Ni}, {Qi}, {Qi},
  {Shen}, {Shi}, {Song}, {Song}, {Su}, {Su}, {Tang}, {Tao}, {Tian}, {Wang},
  {Wang}, {Wang}, {Wang}, {Wang}, {Wang}, {Wang}, {Wang}, {Wang}, {Wang},
  {Wang}, {Wang}, {Wang}, {Wang}, {Wang}, {Wang}, {Wang}, {Wang}, {Wang},
  {Wang}, {Wei}, {Wei}, {Wu}, {Wu}, {Wu}, {Wu}, {Xing}, {Xu}, {Xu}, {Xu},
  {Yan}, {Yang}, {Yang}, {Yang}, {Yang}, {Yao}, {Yu}, {Yuan}, {Yuan}, {Yuan},
  {Yuan}, {Zhai}, {Zhang}, {Zhang}, {Zhang}, {Zhang}, {Zhang}, {Zhang},
  {Zhang}, {Zhang}, {Zhao}, {Zhou}, {Zhou}, {Zhu}, {Zhu}, {Zou}, \&
  {Zuo}}]{LuoEtal15}
{Luo}, A.~L., {Zhao}, Y.-H., {Zhao}, G., {et~al.} 2015, Research in Astronomy
  and Astrophysics, 15, 1095

\bibitem[{{Meisner} {et~al.}(2018){Meisner}, {Lang}, \&
  {Schlegel}}]{MeisnerEtal18}
{Meisner}, A.~M., {Lang}, D., \& {Schlegel}, D.~J. 2018, Research Notes of the
  American Astronomical Society, 2, 1

\bibitem[{{Moffat}(1969)}]{Moffat69}
{Moffat}, A.~F.~J. 1969, \aap, 3, 455

\bibitem[{{Myers} {et~al.}(2003){Myers}, {Jackson}, {Browne}, {de Bruyn},
  {Pearson}, {Readhead}, {Wilkinson}, {Biggs}, {Blandford}, \&
  {Fassnacht}}]{MyersEtal03}
{Myers}, S.~T., {Jackson}, N.~J., {Browne}, I.~W.~A., {et~al.} 2003, \mnras,
  341, 1

\bibitem[{{Nakoneczny} {et~al.}(2019){Nakoneczny}, {Bilicki}, {Solarz},
  {Pollo}, {Maddox}, {Spiniello}, {Brescia}, \& {Napolitano}}]{Nakoneczny2019}
{Nakoneczny}, S., {Bilicki}, M., {Solarz}, A., {et~al.} 2019, \aap, 624, A13

\bibitem[{{Nierenberg} {et~al.}(2017){Nierenberg}, {Treu}, {Brammer}, {Peter},
  {Fassnacht}, {Keeton}, {Kochanek}, {Schmidt}, {Sluse}, \&
  {Wright}}]{NierenbergEtal17}
{Nierenberg}, A.~M., {Treu}, T., {Brammer}, G., {et~al.} 2017, \mnras, 471,
  2224

\bibitem[{{Oguri} {et~al.}(2006){Oguri}, {Inada}, {Pindor}, {Strauss},
  {Richards}, {Hennawi}, {Turner}, {Lupton}, {Schneider}, \&
  {Fukugita}}]{OguriEtal06}
{Oguri}, M., {Inada}, N., {Pindor}, B., {et~al.} 2006, \aj, 132, 999

\bibitem[{{Oguri} {et~al.}(2012){Oguri}, {Inada}, {Strauss}, {Kochanek},
  {Kayo}, {Shin}, {Morokuma}, {Richards}, {Rusu}, \& {Frieman}}]{OguriEtal12}
{Oguri}, M., {Inada}, N., {Strauss}, M.~A., {et~al.} 2012, \aj, 143, 120

\bibitem[{{Oguri} {et~al.}(2008){Oguri}, {Inada}, {Strauss}, {Kochanek},
  {Richards}, {Schneider}, {Becker}, {Fukugita}, {Gregg}, \&
  {Hall}}]{OguriEtal08}
{Oguri}, M., {Inada}, N., {Strauss}, M.~A., {et~al.} 2008, \aj, 135, 512

\bibitem[{{Refsdal}(1964)}]{Refsdal64}
{Refsdal}, S. 1964, \mnras, 128, 307

\bibitem[{{Savary} {et~al.}(2021){Savary}, {Rojas}, {Maus}, {Cl{\'e}ment},
  {Courbin}, {Gavazzi}, {Chan}, {Lemon}, {Vernardos}, {Ca{\~n}ameras},
  {Schuldt}, {Suyu}, {Cuillandre}, {Fabbro}, {Gwyn}, {Hudson}, {Kilbinger},
  {Scott}, \& {Stone}}]{SavaryEtal21}
{Savary}, E., {Rojas}, K., {Maus}, M., {et~al.} 2021, arXiv e-prints,
  arXiv:2110.11972

\bibitem[{{Secrest} {et~al.}(2015){Secrest}, {Dudik}, {Dorland}, {Zacharias},
  {Makarov}, {Fey}, {Frouard}, \& {Finch}}]{SecrestEtal15}
{Secrest}, N.~J., {Dudik}, R.~P., {Dorland}, B.~N., {et~al.} 2015, \apjs, 221,
  12

\bibitem[{{S{\'e}rsic}(1963)}]{Sersic63}
{S{\'e}rsic}, J.~L. 1963, Boletin de la Asociacion Argentina de Astronomia La
  Plata Argentina, 6, 41

\bibitem[{{Shen} {et~al.}(2021){Shen}, {Chen}, {Hwang}, {Liu}, {Zakamska},
  {Oguri}, {Li}, {Lazio}, \& {Breiding}}]{ShenEtal21}
{Shen}, Y., {Chen}, Y.-C., {Hwang}, H.-C., {et~al.} 2021, arXiv e-prints,
  arXiv:2105.03298

\bibitem[{{Shu} {et~al.}(2018){Shu}, {Marques-Chaves}, {Evans}, \&
  {P{\'e}rez-Fournon}}]{ShuEtal18}
{Shu}, Y., {Marques-Chaves}, R., {Evans}, N.~W., \& {P{\'e}rez-Fournon}, I.
  2018, \mnras, 481, L136

\bibitem[{{Sonnenfeld} {et~al.}(2018){Sonnenfeld}, {Chan}, {Shu}, {More},
  {Oguri}, {Suyu}, {Wong}, {Lee}, {Coupon}, {Yonehara}, {Bolton}, {Jaelani},
  {Tanaka}, {Miyazaki}, \& {Komiyama}}]{SonnenfeldEtal18}
{Sonnenfeld}, A., {Chan}, J. H.~H., {Shu}, Y., {et~al.} 2018, \pasj, 70, S29

\bibitem[{{Sonnenfeld} {et~al.}(2019){Sonnenfeld}, {Jaelani}, {Chan}, {More},
  {Suyu}, {Wong}, {Oguri}, \& {Lee}}]{SonnenfeldEtal19}
{Sonnenfeld}, A., {Jaelani}, A.~T., {Chan}, J. H.~H., {et~al.} 2019, arXiv
  e-prints, arXiv:1904.10465

\bibitem[{{Sonnenfeld} {et~al.}(2020){Sonnenfeld}, {Verma}, {More}, {Baeten},
  {Macmillan}, {Wong}, {Chan}, {Jaelani}, {Lee}, {Oguri}, {Rusu}, {Veldthuis},
  {Trouille}, {Marshall}, {Hutchings}, {Allen}, {O'Donnell}, {Cornen}, {Davis},
  {McMaster}, {Lintott}, \& {Miller}}]{SonnenfeldEtal20}
{Sonnenfeld}, A., {Verma}, A., {More}, A., {et~al.} 2020, \aap, 642, A148

\bibitem[{{Stern} \& {Walton}(2020)}]{Stern&Walton20}
{Stern}, D. \& {Walton}, D.~J. 2020, \apjl, 895, L38

\bibitem[{{Stetson}(1987)}]{Stetson87}
{Stetson}, P.~B. 1987, \pasp, 99, 191

\bibitem[{{Suyu} {et~al.}(2012){Suyu}, {Hensel}, {McKean}, {Fassnacht}, {Treu},
  {Halkola}, {Norbury}, {Jackson}, {Schneider}, {Thompson}, {Auger},
  {Koopmans}, \& {Matthews}}]{SuyuEtal12}
{Suyu}, S.~H., {Hensel}, S.~W., {McKean}, J.~P., {et~al.} 2012, \apj, 750, 10

\bibitem[{{Vegetti} {et~al.}(2012){Vegetti}, {Lagattuta}, {McKean}, {Auger},
  {Fassnacht}, \& {Koopmans}}]{VegettiEtal12}
{Vegetti}, S., {Lagattuta}, D.~J., {McKean}, J.~P., {et~al.} 2012, \nat, 481,
  341

\bibitem[{{Verde} {et~al.}(2019){Verde}, {Treu}, \& {Riess}}]{VerdeEtal19}
{Verde}, L., {Treu}, T., \& {Riess}, A.~G. 2019, Nature Astronomy, 3, 891

\bibitem[{{Wong} {et~al.}(2018){Wong}, {Sonnenfeld}, {Chan}, {Rusu}, {Tanaka},
  {Jaelani}, {Lee}, {More}, {Oguri}, {Suyu}, \& {Komiyama}}]{WongEtal18}
{Wong}, K.~C., {Sonnenfeld}, A., {Chan}, J. H.~H., {et~al.} 2018, \apj, 867,
  107

\bibitem[{{Wong} {et~al.}(2019){Wong}, {Suyu}, {Chen}, {Rusu}, {Millon},
  {Sluse}, {Bonvin}, {Fassnacht}, {Taubenberger}, {Auger}, {Birrer}, {Chan},
  {Courbin}, {Hilbert}, {Tihhonova}, {Treu}, {Agnello}, {Ding}, {Jee},
  {Komatsu}, {Shajib}, {Sonnenfeld}, {Bland ford}, {Koopmans}, {Marshall}, \&
  {Meylan}}]{WongEtal19}
{Wong}, K.~C., {Suyu}, S.~H., {Chen}, G. C.~F., {et~al.} 2019, arXiv e-prints,
  arXiv:1907.04869

\end{thebibliography}

\appendix
\section{Test for known lenses}
\label{sec:known}

\begin{figure}[h]
\centering
\includegraphics[scale=0.35]{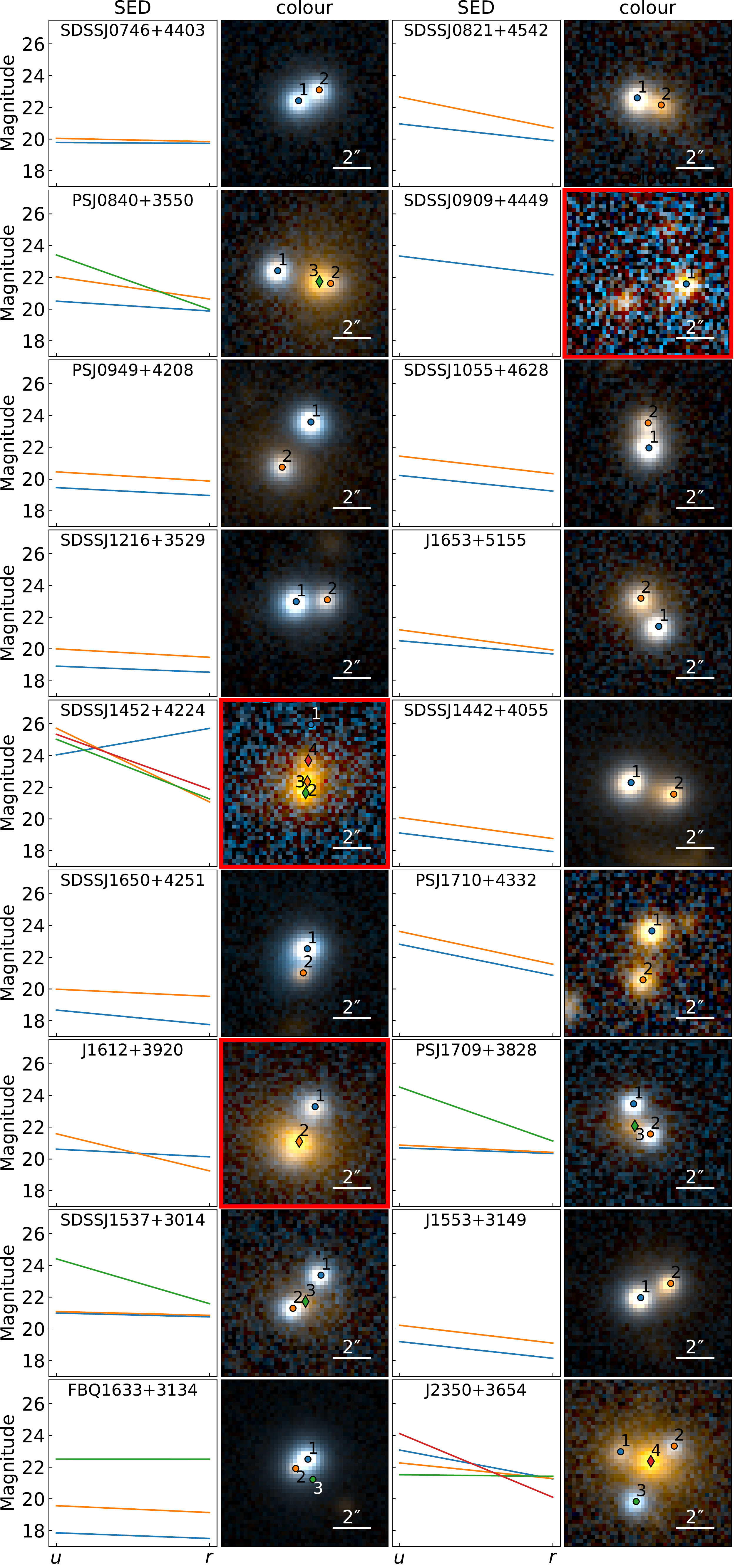}
\vskip -10pt
\caption{
Known lens test for the classification method in \sref{sec:method}. The labels are same as \fref{fig:chitah} (a) and (b). The lenses that are not able to be recovered are highlighted with the red frames. SDSSJ0909+4449 has an overly wide image separation $\sim18\arcsec$. SDSSJ1452+4224 has bad pixels in CFIS-$u$ image 1. J1612+3920 has a faint counter image hidden to the bottom-left of the lensing galaxy.
}
\label{fig:known}
\end{figure}

There are 18 optically bright known lensed quasars with CFIS DR2 imaging. We recover 15 of them via the number of blue source detections $2\leq\nimg\leq5$; see \fref{fig:known} and \tref{tab:known}. The three failures are listed below:
\begin{enumerate}
    \item SDSSJ0909$+$4449: overly wide separation \citep[around $18\arcsec$,][]{ShuEtal18};
    \item SDSSJ1452$+$4224: bad pixels in CFIS-$u$;
    \item J1612$+$3920: omission of one lensed image identification.
\end{enumerate}

\begin{table}[h]
  \caption{
  Known lenses for testing the classification method in \sref{sec:method}.}
  \centering
  \begin{tabular}{rrrc}
  \hline
  \hline
  \multicolumn{1}{c}{Name} & \multicolumn{1}{c}{R.A. [$\deg$]} & \multicolumn{1}{c}{Dec. [$\deg$]} & $\nimg$ \\
  \hline
                SDSSJ0746$+$4403 & $   116.721$ & $    44.064$ & $  2$ \\
                SDSSJ0821$+$4542 & $   125.494$ & $    45.712$ & $  2$ \\
                  PSJ0840$+$3550 & $   130.138$ & $    35.833$ & $  2$ \\
      SDSSJ0909$+$4449$^\dagger$ & $   137.440$ & $    44.832$ & $  1$ \\
                  PSJ0949$+$4208 & $   147.478$ & $    42.134$ & $  2$ \\
                SDSSJ1055$+$4628 & $   163.939$ & $    46.478$ & $  2$ \\
                SDSSJ1216$+$3529 & $   184.192$ & $    35.495$ & $  2$ \\
                    J1653$+$5155 & $   253.439$ & $    51.918$ & $  2$ \\
      SDSSJ1452$+$4224$^\dagger$ & $   223.048$ & $    42.408$ & $  1$ \\
                SDSSJ1442$+$4055 & $   220.728$ & $    40.927$ & $  2$ \\
                SDSSJ1650$+$4251 & $   252.681$ & $    42.864$ & $  2$ \\
                  PSJ1710$+$4332 & $   257.743$ & $    43.543$ & $  2$ \\
          J1612$+$3920$^\dagger$ & $   243.051$ & $    39.347$ & $  1$ \\
                  PSJ1709$+$3828 & $   257.370$ & $    38.467$ & $  2$ \\
                SDSSJ1537$+$3014 & $   234.393$ & $    30.248$ & $  2$ \\
                    J1553$+$3149 & $   238.409$ & $    31.825$ & $  2$ \\
                  FBQ1633$+$3134 & $   248.454$ & $    31.570$ & $  3$ \\
                    J2350$+$3654 & $   357.531$ & $    36.910$ & $  3$ \\
  \hline
  \end{tabular}
  \tablefoot{The number of blue point source detections is denoted as $\nimg$. We highlight with $\dagger$ the lenses that are not recovered.
  }
  \label{tab:known}
\end{table}

\section{Result from light modelling}
\label{sec:photometry}

We measure the CFIS astrometry and photometry from light modelling in \tref{tab:cand_mag}. See details in \sref{sec:light}. These candidates are the same as in \fref{fig:cand_G8NOT} and \tref{tab:cand_G8NOT}.

\section{Two unresolved spectra}
\label{sec:spec_other}
\begin{figure*}[t!]
\centering
\includegraphics[scale=0.45]{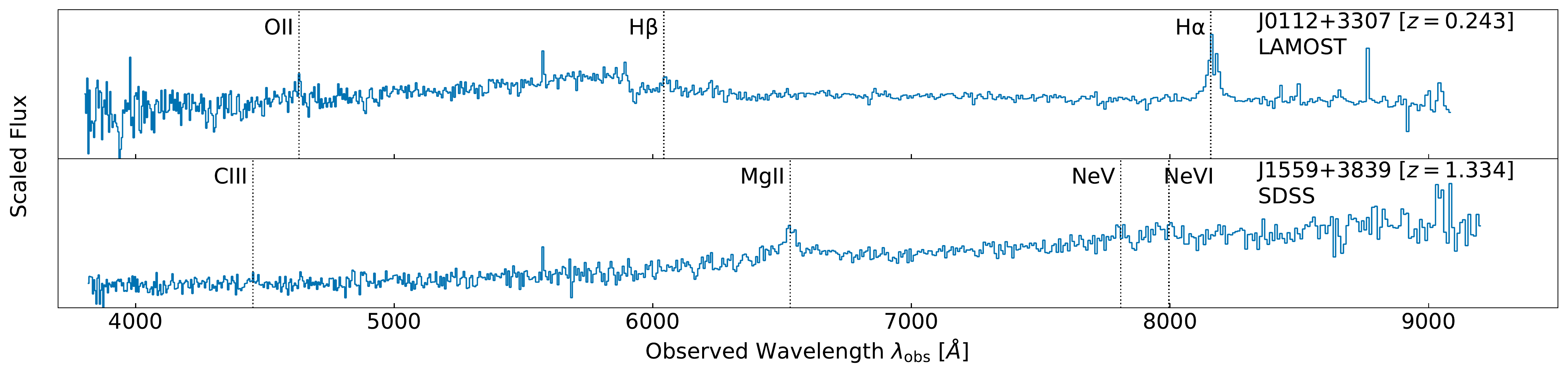}
\caption{One-dimensional spectra of J0112+3307 from LAMOST and J1559+3839 from SDSS.
}
\label{fig:spec_others}
\end{figure*}

We measure the redshifts of two additional systems: J0112+3307 and J1559+3839, from the public spectra of LAMOST and SDSS, respectively; see \fref{fig:spec_others}. 
\clearpage
\onecolumn
\begin{longtable}{ccrrrr}
\label{tab:cand_mag}\\
\caption{Result of light modelling for the candidates in \fref{fig:cand_G8NOT}.}
\\
\hline\hline
\multicolumn{1}{c}{Name} & \multicolumn{1}{c}{Component} & \multicolumn{1}{c}{$\Delta\alpha$ [$\arcsec$]} & \multicolumn{1}{c}{$\Delta\delta$ [$\arcsec$]} & CFIS-$u$ & CFIS-$r$ \\
\hline
\endfirsthead
\caption{continued.}\\
\hline\hline
\multicolumn{1}{c}{Name} & \multicolumn{1}{c}{Component} & \multicolumn{1}{c}{$\Delta\alpha$ [$\arcsec$]} & \multicolumn{1}{c}{$\Delta\delta$ [$\arcsec$]} & CFIS-$u$ & CFIS-$r$ \\
\hline
\endhead
\hline
\endfoot
UNIONS           J0112+3307 & $A$ & $     0.077$ & $    -0.074$ & $     21.29$ & $     19.27$\\
                          & $B$ & $    -0.682$ & $    -0.779$ & $     21.21$ & $     20.21$\\
                          & $G$ & $    -0.552$ & $    -0.605$ & $     20.64$ & $     18.35$\\
\hline
UNIONS           J0800+4059 & $A$ & $    -0.007$ & $     0.160$ & $     20.82$ & $     19.94$\\
                          & $G$ & $    -0.779$ & $    -0.135$ & $     23.06$ & $     20.52$\\
\hline
UNIONS           J0805+3550 & $A$ & $    -1.082$ & $     0.027$ & $     19.47$ & $     19.23$\\
                          & $B$ & $     0.059$ & $     0.089$ & $     20.38$ & $     19.60$\\
\hline
UNIONS           J0840+4631 & $A$ & $     0.039$ & $     0.097$ & $     20.46$ & $     19.63$\\
                          & $G$ & $    -2.050$ & $    -1.223$ & $     20.96$ & $     18.79$\\
\hline
UNIONS           J0922+4028 & $A$ & $    -0.264$ & $     0.893$ & $     22.55$ & $     21.49$\\
                          & $B$ & $    -1.588$ & $     0.626$ & $     23.11$ & $     22.23$\\
\hline
UNIONS           J0926+3059 & $A$ & $    -0.096$ & $     0.030$ & $     20.02$ & $     19.90$\\
                          & $B$ & $     0.821$ & $    -1.822$ & $     21.31$ & $     21.13$\\
\hline
UNIONS           J0938+4216 & $A$ & $     0.034$ & $     0.019$ & $     20.02$ & $     19.60$\\
                          & $B$ & $    -0.596$ & $    -0.470$ & $     21.03$ & $     19.98$\\
\hline
UNIONS           J0943+3532 & $A$ & $    -0.583$ & $     0.634$ & $     19.70$ & $     20.02$\\
                          & $B$ & $    -0.066$ & $    -0.017$ & $     19.92$ & $     20.84$\\
                          & $G$ & $    -0.263$ & $     0.263$ & $     21.45$ & $     19.44$\\
\hline
UNIONS           J0958+4325 & $A$ & $     0.099$ & $     0.374$ & $     20.40$ & $     20.31$\\
                          & $B$ & $    -0.782$ & $    -0.065$ & $     22.38$ & $     21.25$\\
\hline
UNIONS           J1006+3040 & $A$ & $    -0.881$ & $     0.193$ & $     19.54$ & $     19.58$\\
                          & $B$ & $    -0.062$ & $     0.020$ & $     20.37$ & $     20.43$\\
\hline
UNIONS           J1039+5126 & $A$ & $    -0.022$ & $    -0.045$ & $     20.20$ & $     19.96$\\
                          & $B$ & $    -1.029$ & $     1.316$ & $     20.43$ & $     20.04$\\
\hline
UNIONS           J1111+3804 & $A$ & $    -0.002$ & $    -0.034$ & $     22.45$ & $     20.85$\\
                          & $B$ & $    -0.339$ & $     1.901$ & $     23.02$ & $     21.76$\\
                          & $G$ & $     0.094$ & $     1.532$ & $     25.47$ & $     22.11$\\
\hline
UNIONS           J1136+4227 & $A$ & $    -0.623$ & $    -2.083$ & $     19.09$ & $     19.03$\\
                          & $B$ & $     0.062$ & $    -0.053$ & $     21.12$ & $     21.07$\\
\hline
UNIONS           J1146+3254 & $A$ & $    -0.022$ & $     0.037$ & $     20.61$ & $     20.75$\\
                          & $B$ & $    -1.761$ & $    -1.676$ & $     22.43$ & $     22.85$\\
                          & $G$ & $    -1.561$ & $    -1.402$ & $     23.51$ & $     19.97$\\
\hline
UNIONS           J1201+4013 & $A$ & $    -0.060$ & $    -0.019$ & $     19.16$ & $     18.92$\\
                          & $B$ & $    -0.744$ & $    -1.476$ & $     21.53$ & $     20.99$\\
                          & $G$ & $    -0.597$ & $    -1.169$ & $     21.51$ & $     20.53$\\
\hline
UNIONS           J1219+3453 & $A$ & $     0.014$ & $    -4.121$ & $     21.67$ & $     21.20$\\
                          & $B$ & $     0.057$ & $     0.029$ & $     21.26$ & $     20.65$\\
                          & $C$ & $     0.126$ & $    -3.001$ & $     24.54$ & $     20.98$\\
\hline
UNIONS           J1223+3103 & $A$ & $    -0.023$ & $     0.053$ & $     19.78$ & $     19.18$\\
                          & $B$ & $    -0.639$ & $    -0.749$ & $     22.26$ & $     21.16$\\
\hline
UNIONS           J1325+3629 & $A$ & $    -0.470$ & $     0.438$ & $     20.04$ & $     19.80$\\
                          & $B$ & $     0.054$ & $    -0.005$ & $     21.72$ & $     21.30$\\
\hline
UNIONS           J1454+4146 & $A$ & $    -0.004$ & $    -0.099$ & $     20.66$ & $     19.99$\\
                          & $B$ & $    -0.943$ & $    -1.156$ & $     21.58$ & $     21.06$\\
\hline
UNIONS           J1529+3841 & $A$ & $    -0.957$ & $    -0.003$ & $     19.10$ & $     18.70$\\
                          & $B$ & $    -0.042$ & $    -0.068$ & $     21.89$ & $     20.93$\\
\hline
UNIONS           J1537+3223 & $A$ & $    -1.033$ & $     0.270$ & $     19.71$ & $     19.90$\\
                          & $B$ & $    -0.086$ & $     0.034$ & $     20.12$ & $     19.97$\\
\hline
UNIONS           J1551+3423 & $A$ & $     0.038$ & $     0.088$ & $     19.64$ & $     19.24$\\
                          & $B$ & $    -0.741$ & $    -0.811$ & $     22.01$ & $     21.36$\\
\hline
UNIONS           J1555+3118 & $A$ & $    -1.459$ & $    -0.914$ & $     22.87$ & $     18.97$\\
                          & $B$ & $    -0.099$ & $     0.004$ & $     24.27$ & $     20.18$\\
                          & $G$ & $    -0.995$ & $    -0.487$ & $     30.00$ & $     20.64$\\
\hline
UNIONS           J1555+3703 & $A$ & $    -0.096$ & $     0.023$ & $     19.63$ & $     19.48$\\
                          & $B$ & $    -1.021$ & $     0.080$ & $     21.60$ & $     21.18$\\
\hline
UNIONS           J1555+3813 & $A$ & $     0.026$ & $    -0.080$ & $     21.57$ & $     20.70$\\
                          & $B$ & $    -0.608$ & $    -0.478$ & $     21.87$ & $     21.42$\\
\hline
UNIONS           J1559+3839 & $A$ & $     0.153$ & $     0.245$ & $     23.60$ & $     22.07$\\
                          & $B$ & $    -0.292$ & $    -0.520$ & $     24.04$ & $     22.99$\\
                          & $G$ & $     0.057$ & $    -0.065$ & $     30.00$ & $     20.33$\\
\hline
UNIONS           J1633+4556 & $A$ & $    -0.102$ & $     0.053$ & $     19.69$ & $     19.22$\\
                          & $B$ & $    -0.615$ & $    -0.593$ & $     22.25$ & $     21.15$\\
\hline
UNIONS           J1655+3746 & $A$ & $    -0.001$ & $    -0.161$ & $     20.56$ & $     20.37$\\
                          & $B$ & $     0.964$ & $     0.241$ & $     21.59$ & $     21.18$\\
\hline
UNIONS           J1739+4355 & $A$ & $     0.066$ & $    -0.683$ & $     22.14$ & $     20.73$\\
                          & $B$ & $     0.247$ & $     0.844$ & $     22.03$ & $     21.01$\\
\hline
UNIONS           J2335+3201 & $A$ & $    -0.152$ & $    -0.076$ & $     19.94$ & $     19.34$\\
                          & $B$ & $    -0.185$ & $     0.543$ & $     21.31$ & $     20.48$\\
\end{longtable}
\tablefoot{The astrometric positions ($\Delta\alpha$ and $\Delta\delta$) are measured with respect to the coordinates in \tref{tab:cand_G8NOT}.
The photometric magnitudes are in the AB system.
}


\end{document}